\documentclass[conference]{IEEEtran}
\IEEEoverridecommandlockouts
\usepackage{cite}
\usepackage{amsmath,amssymb,amsfonts}
\usepackage{algorithmic}
\usepackage{graphicx}
\usepackage{textcomp}
\usepackage{xcolor}
\def\BibTeX{{\rm B\kern-.05em{\sc i\kern-.025em b}\kern-.08em
    T\kern-.1667em\lower.7ex\hbox{E}\kern-.125emX}}

\usepackage{graphicx}
\usepackage{color}

\newcommand{\eg}{e.g.~}
\newcommand{\ie}{i.e.~}

\newcommand{\etal}{et al.~}
\newcommand{\code}[1]{\texttt{#1}}
\newcommand{\confuzzius}{\textsc{ConFuzzius}}

\usepackage{array}
\newcolumntype{L}[1]{>{\raggedright\let\newline\\\arraybackslash\hspace{0pt}}m{#1}}
\newcolumntype{C}[1]{>{\centering\let\newline\\\arraybackslash\hspace{0pt}}m{#1}}
\newcolumntype{R}[1]{>{\raggedleft\let\newline\\\arraybackslash\hspace{0pt}}m{#1}}

\usepackage{multirow}
\usepackage{float}
\usepackage{algorithm}
\usepackage{algorithmic}
\usepackage{url}
\usepackage{pifont}
\newcommand{\cmark}{\ding{51}}
\newcommand{\xmark}{\ding{55}}
\usepackage{wasysym}

\usepackage{tikz}
\usetikzlibrary{positioning,arrows,tikzmark,calc}

\usepackage{listings, xcolor}

\definecolor{verylightgray}{rgb}{.97,.97,.97}

\lstdefinelanguage{Solidity}{
	keywords=[1]{anonymous, assembly, assert, balance, break, call, callcode, case, catch, class, constant, continue, contract, debugger, default, delegatecall, delete, do, else, emit, event, export, external, false, finally, for, function, constructor, gas, if, implements, import, in, indexed, instanceof, interface, internal, is, length, library, log0, log1, log2, log3, log4, memory, modifier, new, payable, pragma, private, protected, public, pure, push, require, return, returns, revert, selfdestruct, send, storage, struct, suicide, super, switch, then, this, throw, transfer, true, try, typeof, using, value, view, while, with, addmod, ecrecover, keccak256, mulmod, ripemd160, sha256, sha3}, %
	keywordstyle=[1]\color{blue}\bfseries,
	keywords=[2]{address, bool, byte, bytes, bytes1, bytes2, bytes3, bytes4, bytes5, bytes6, bytes7, bytes8, bytes9, bytes10, bytes11, bytes12, bytes13, bytes14, bytes15, bytes16, bytes17, bytes18, bytes19, bytes20, bytes21, bytes22, bytes23, bytes24, bytes25, bytes26, bytes27, bytes28, bytes29, bytes30, bytes31, bytes32, enum, int, int8, int16, int24, int32, int40, int48, int56, int64, int72, int80, int88, int96, int104, int112, int120, int128, int136, int144, int152, int160, int168, int176, int184, int192, int200, int208, int216, int224, int232, int240, int248, int256, mapping, string, uint, uint8, uint16, uint24, uint32, uint40, uint48, uint56, uint64, uint72, uint80, uint88, uint96, uint104, uint112, uint120, uint128, uint136, uint144, uint152, uint160, uint168, uint176, uint184, uint192, uint200, uint208, uint216, uint224, uint232, uint240, uint248, uint256, var, void, ether, finney, szabo, wei, days, hours, minutes, seconds, weeks, years},	%
	keywordstyle=[2]\color{teal}\bfseries,
	keywords=[3]{block, blockhash, coinbase, difficulty, gaslimit, number, timestamp, msg, data, gas, sender, sig, value, now, tx, gasprice, origin},	%
	keywordstyle=[3]\color{violet}\bfseries,
	identifierstyle=\color{black},
	sensitive=false,
	comment=[l]{//},
	morecomment=[s]{/*}{*/},
	commentstyle=\color{gray}\ttfamily,
	stringstyle=\color{red}\ttfamily,
	morestring=[b]',
	morestring=[b]"
}

\usepackage{balance}
\usepackage{adjustbox}

\begin{document}

\title{\textsc{ConFuzzius:} A Data Dependency-Aware Hybrid Fuzzer for Smart Contracts}

\author{
\IEEEauthorblockN{
Christof Ferreira Torres\IEEEauthorrefmark{1}, 
Antonio Ken Iannillo\IEEEauthorrefmark{1}, 
Arthur Gervais\IEEEauthorrefmark{2},
and 
Radu State\IEEEauthorrefmark{1}
}
\IEEEauthorblockA{
\IEEEauthorrefmark{1}SnT, University of Luxembourg\\
}
\IEEEauthorblockA{
\IEEEauthorrefmark{2}Imperial College London\\
}
}

\maketitle

\begin{abstract}
Smart contracts are Turing-complete programs that are executed across a blockchain.
Unlike traditional programs, once deployed, they cannot be modified. 
As smart contracts carry more value, they become more of an exciting target for attackers. 
Over the last years, they suffered from exploits costing millions of dollars due to simple programming mistakes.
As a result, a variety of tools for detecting bugs have been proposed.
Most of these tools rely on symbolic execution, which may yield false positives due to over-approximation.
Recently, many fuzzers have been proposed to detect bugs in smart contracts.
However, these tend to be more effective in finding shallow bugs and less effective in finding bugs that lie deep in the execution, therefore achieving low code coverage and many false negatives. An alternative that has proven to achieve good results in traditional programs is hybrid fuzzing, a combination of symbolic execution and fuzzing. 

In this work, we study hybrid fuzzing on smart contracts and present \confuzzius{}, the first hybrid fuzzer for smart contracts. 
\confuzzius{} uses evolutionary fuzzing to exercise shallow parts of a smart contract and constraint solving to generate inputs that satisfy complex conditions that prevent evolutionary fuzzing from exploring deeper parts.
Moreover, \confuzzius{} leverages dynamic data dependency analysis to efficiently generate sequences of transactions that are more likely to result in contract states in which bugs may be hidden.
We evaluate the effectiveness of \confuzzius{} by comparing it with state-of-the-art symbolic execution tools and fuzzers for smart contracts.
Our evaluation on a curated dataset of 128 contracts and a dataset of 21K real-world contracts shows that our hybrid approach detects more bugs than state-of-the-art tools (up to 23\%)
and that it outperforms existing tools in terms of code coverage (up to 69\%).
We also demonstrate that data dependency analysis can boost bug detection up to 18\%.
\end{abstract}


\section{Introduction}
\label{sec:introduction}

The inception of immutable, blockchain-based smart contracts has shown how to enable multiple mistrusting parties to trade and interact without relying on a centralized, trusted third party. 
The immutability of a contract is crucial: if at least one of the engaging parties were allowed to modify a digital contract, the contract's trust would vanish. 
Unlike traditional legal contracts, smart contracts do not allow a dispute resolution with a neutral third party. 
Most importantly, smart contracts cannot be nullified ---  parties cannot revoke any deployed smart contract, even if its code figures undeniable software bugs.
Therefore, this very immutability comes at a price: smart contracts must be tested extensively before exposing them and their users to significant monetary value. 
In the past, simple vulnerabilities (\eg missing access control~\cite{petrov2017another}) and subtle vulnerabilities (\eg reentrancy~\cite{siegel2016understanding}) have led to losses exceeding many tens of millions of USD.

We can verify the behavior of a smart contract with four different approaches.
(i) \emph{Unit testing} requires manual effort to cover the different sections of the code, but it unveils only a limited number of bugs within the test cases. 
(ii) \emph{Symbolic execution} analyzes contract behavior abstractly but performs slowly on complex contracts (path explosion problem). 
(iii) \emph{Static analysis} does not execute code and over-approximates the contract behavior --- it can capture the entire contract execution surface, but it exhibits false positives that must be manually inspected. 
Finally, (iv) \emph{fuzzing} tests a contract reasonably fast by automatically generating test cases, with a generally lower false positive rate than static analysis. Fuzzing, however, can suffer from low code coverage, especially when input is fuzzed at random and hence does not overcome simple input sanity verification.

When fuzzing smart contracts, we face the following three challenges:
1) \emph{input generation}, 2) \emph{stateful exploration}, and 3) \emph{environmental dependencies}.
When it comes to input generation, the input space can be significantly broad. However, the solution might be limited to a specific point. For example, if a condition requires an input value of type \code{uint256} to equate to $42$, then the probability of randomly generating $42$ as input is tremendously small.
Moreover, smart contracts are stateful applications, \ie the execution may depend on a state that is only achievable following a specific sequence of inputs.
Finally, the runtime environment of smart contracts exposes additional inputs related to the underlying blockchain protocol, such as the current block timestamp or other contracts deployed on the blockchain. 
As a result, the execution flow of smart contracts may depend on environmental information besides transactional information.

We solve these three challenges as follows. 
In tandem with the fuzzing procedure, we employ symbolic taint analysis to generate path constraints on tainted inputs. 
Once we detect that the fuzzer is not progressing, we activate a constraint solver to solve the constraint in question. We collect this solution within a mutation pool, from which the fuzzer can draw to move past the challenging contract condition. 
Existing hybrid fuzzing approaches, \eg Driller~\cite{stephens2016driller}, cease the fuzzer when they are stuck and switch to concolic execution to get past the complex condition. Then, they restart the fuzzer once passed the condition. 
Our approach keeps the fuzzer running and only uses constraint solving to generate inputs on the fly, which will eventually be picked by the fuzzer via the mutation pools.
In addition to constraint solving, we perform a path termination analysis to purge irrelevant inputs from the mutation pools.
To deal with the statefulness of smart contracts, we chose to take advantage of the selection and crossover operators of genetic algorithms.
Genetic algorithms follow three main steps: selection, crossover, and mutation. 
The selection operator's task is to choose two individuals from the population, which are afterwards combined by the crossover operator to create two new individuals.
The challenge here is to generate meaningful combinations of inputs.
Therefore, data dependencies between individuals guide our selection and crossover operators that accept two individuals only if they follow a \emph{read-after-write} (RAW) data dependency.
Finally, to solve the third and last challenge, we instrument the execution environment (\ie the Ethereum Virtual Machine) to fuzz environmental information and model the input to a contract as a tuple consisting of transactional \emph{and} environmental data.

\noindent
\newline
\textbf{Contributions.} 
Our main contributions are as follows:
\begin{itemize}
    \item To the best of our knowledge, we propose the first design of a hybrid fuzzer for smart contracts.
    \item We present a novel method to efficiently create meaningful sequences of inputs at runtime by leveraging dynamic data dependencies between state variables.
    \item We introduce \confuzzius{}, the first implementation of a hybrid fuzzer for smart contracts. 
    \item We evaluate \confuzzius{} on a set of 128 curated smart contracts as well as 21K real-world smart contracts, and demonstrate that our approach not only detects more vulnerabilities (up to 23\%) but also achieves more code coverage (up to 69\%) than existing symbolic execution tools and fuzzers.
\end{itemize}

\section{Background}
\label{sec:background}

This section provides the required background on Ethereum smart contracts and fuzzing in order to better understand the approach proposed in this work.

\subsection{Ethereum Smart Contracts}

\noindent
\textbf{Smart Contracts.}
Ethereum~\cite{wood2014ethereum} enables the execution of so-called \emph{smart contracts}. 
These are fully-fledged programs stored and executed across the Ethereum blockchain, a network of mutually distrusting nodes.
Ethereum supports two types of accounts, externally owned accounts (\ie user accounts) and contract accounts (\ie smart contracts).
Smart contracts are different from traditional programs in many ways.
They own a balance and are identifiable via a 160-bit address.
They are developed using a dedicated high-level programming language, such as Solidity~\cite{solidity}, that compiles into low-level bytecode. This bytecode gets interpreted by the Ethereum Virtual Machine.
By default, smart contracts cannot be removed or updated once deployed. 
It is the task of the developer to implement these capabilities before deployment.
The deployment of smart contracts and the execution of smart contract functions occurs via transactions. 
The data field of a transaction includes both, the name of the function to be executed and its arguments.
Transactions are created by user accounts and afterwards broadcast to the network.
They contain a sender and a recipient. The latter can be the address of a user account or a contract account.
Besides carrying data, transactions may also carry a monetary value in the form of ether (Ethereum's own cryptocurrency).

\noindent
\newline
\textbf{Ethereum Virtual Machine.} 
The Ethereum Virtual Machine (EVM) is a purely stack-based, register-less virtual machine that supports a Turing-complete set of instructions.
Although the instruction set allows for Turing-complete programs, the instructions' capabilities are limited to the sole manipulation of the blockchain's state.
The instruction set provides a variety of operations, ranging from generic operations, such as arithmetic operations or control-flow statements, to more specific ones, such as the modification of a contract's storage or the querying of properties related to the transaction (\eg sender) or the current blockchain state (\eg block number).
Ethereum uses a \emph{gas} mechanism to assure the termination of contracts and prevent denial-of-service attacks.
The gas mechanism associates costs to the execution of every single instruction.
When issuing a transaction, the sender specifies how much gas they are willing to spend to execute the smart contract. 
This amount is known as the \emph{gas limit}.

\subsection{Fuzzing}

\noindent
\textbf{Evolutionary Fuzzing.}
Fuzzing, or fuzz testing, is an automated software testing technique that finds vulnerabilities in programs by feeding malformed or unexpected data as input to programs, executing them, and monitoring the effects.
Evolutionary fuzzing aims at converging towards the discovery of vulnerabilities by using a genetic algorithm (GA). %
A generation of test cases is defined as a \emph{population}, whereas a single test case is an \emph{individual}.
In short, every individual of a generation is evaluated based on a fitness function.
At the end of each generation, solely the fittest individuals are allowed to breed, following Darwin's idea of natural selection, or "survival of the fittest".
Eventually, the individuals will trigger vulnerabilities while converging towards an optimal solution.
We briefly describe the main steps of a GA (see Algorithm~\ref{alg:genetic_algorithm}). 
We start by creating an initial population of individuals, either generated at random or seeded via heuristics, and compute their fitness values (\emph{line 1}).
Based on the fitness value, we select two individuals from the current population, which act as parents for breading (\textit{line 5}). 
Then, we apply crossover and mutation operators on the parents to generate two new individuals, also denoted as offsprings (\textit{lines 6-7}). 
The generation of new individuals continues until the new population reaches the same size as the current one (\textit{line 4}). 
Finally, the new population's fitness values are computed, and we replace the current population with the new population (\textit{lines 9-10}).
This entire process is repeated until a termination condition is met (\textit{line 3}), \eg the maximum number of generations is reached or a maximum amount of time has passed.

\begin{algorithm}
\small
\caption{Pseudo-Code of a Genetic Algorithm}
\label{alg:genetic_algorithm}
\begin{algorithmic}[1]
    \STATE Create initial population and compute its fitness
    \STATE Set initial population as current population
    \WHILE{termination condition is not met}
        \WHILE{new population $<$ current population}
                \STATE \textit{Select} two parents from current population
                \STATE \textit{Recombine} parents to create two new offsprings
                \STATE \textit{Mutate} offsprings and add them to new population
        \ENDWHILE
        \STATE Compute fitness of new population
        \STATE Replace current population with new population
        \STATE Create a new empty population
    \ENDWHILE
\end{algorithmic}
\end{algorithm}

\noindent
\newline
\textbf{Hybrid Fuzzing.}
Although fuzzing is one of the most effective approaches to find vulnerabilities, it often has difficulties in getting past complex path conditions, resulting in low code coverage.
A popular alternative to fuzzing is symbolic execution.
It works by abstractly executing a program and supplying abstract symbols rather than actual (\emph{concrete}) values. 
The execution will then generate symbolic formulas over the input symbols, which can be solved by a constraint solver to prove satisfiability and produce concrete values.
In theory, symbolic execution is capable of discovering and exploring all potential paths in a program.
However, in practice, symbolic execution is often not scalable since the number of explorable paths becomes exponential in more extensive programs (\emph{path explosion problem}).
Another limitation of symbolic execution is the limited capability to interact with the execution environment. 
Programs often interact with their environment by performing calls to libraries, for example. 
Correctly modeling these calls and other environmental information is extremely challenging.
The goal of hybrid fuzzing, or hybrid fuzz testing, is to take advantage of both worlds.
Hybrid fuzzing starts by performing traditional fuzzing until it saturates, \emph{i.e.}, the fuzzer is not capable of covering any new code after running some predetermined number of steps. 
Hybrid fuzzing then automatically switches to symbolic execution to perform an exhaustive search for uncovered branching conditions. 
As soon as the symbolic execution finds an uncovered branching condition, it solves it, and the hybrid fuzzer reverts to fuzzing. 
The interleaving of fuzzing and symbolic execution counts on shallow program paths' quick execution via fuzzing and the execution of complex program paths via symbolic execution.
\section{Overview}
\label{sec:approach}

This section discusses the three main challenges of fuzzing smart contracts via a motivating example and presents our solution towards solving these challenges. 

\subsection{Motivating Example}

\begin{figure}
\begin{lstlisting}[language=Solidity,escapechar=\%]
interface Token {
  function transferFrom(address sender, address recipient, uint256 amount) external returns (bool);
  function allowance(address owner, address spender) external view returns (uint256);
}

contract TokenSale {
  uint256 start = now;
  uint256 end = now + 30 days;
  address wallet = 0xcafebabe...;
  Token token = Token(0x12345678...);
  
  address owner;
  bool sold;

  function Tokensale() public {
    %\tikzmark{a}%%\makebox[0pt][l]{\color{gray!30}\rule[-0.45em]{3.2cm}{1.5em}}%owner = msg.sender;
  }

  function buy() public payable {
    %\makebox[0pt][l]{\color{blue!20}\rule[-0.45em]{3.2cm}{1.5em}}%require(now < end);
    %\makebox[0pt][l]{\color{red!20}\rule[-0.45em]{7.6cm}{1.5em}}%require(msg.value == 42 ether + (now - start) %\makebox[0pt][l]{\color{red!20}\rule[-0.45em]{4.4cm}{1.5em}}%/ 60 / 60 / 24 * 1 ether);
    %\makebox[0pt][l]{\color{blue!20}\rule[-0.45em]{7.4cm}{1.5em}}%require(token.transferFrom(this, msg.sender, %\makebox[0pt][l]{\color{blue!20}\rule[-0.45em]{5.4cm}{1.5em}}%token.allowance(wallet, this)));
    %\makebox[0pt][l]{\color{gray!30}\rule[-0.45em]{2.0cm}{1.5em}}%sold = true;%\tikzmark{c}%
  }

  function withdraw() public {
    %\tikzmark{b}%%\makebox[0pt][l]{\color{gray!30}\rule[-0.45em]{4.9cm}{1.5em}}%require(msg.sender == owner);
    %\makebox[0pt][l]{\color{blue!20}\rule[-0.45em]{3.4cm}{1.5em}}%require(now >= end);
    %\makebox[0pt][l]{\color{gray!30}\rule[-0.45em]{2.4cm}{1.5em}}%require(sold);%\tikzmark{d}%
    owner.transfer(address(this).balance);
  }
}
%
\begin{tikzpicture}[overlay, remember picture]
\begin{scope}[transform canvas={yshift=-2pt}]
\draw[->] ([xshift=-3pt, yshift=3pt]pic cs:a) -| ([xshift=-15pt,yshift=3pt]pic cs:b) -- +(12pt,0);
\draw[->] ([xshift=2pt, yshift=3pt]pic cs:c) -| ([xshift=80pt,yshift=3pt]pic cs:d) -- +(-77pt,0);
\end{scope}
\end{tikzpicture}
%
\end{lstlisting}
\caption{Example of a vulnerable token sale smart contract. Lines highlighted in red represent complex conditions, whereas lines highlighted in gray illustrate read-after-write data dependencies and finally, lines highlighted in blue depict environmental dependencies.}
\label{fig:example}
\end{figure}

Suppose a user participated in an initial coin offering (ICO) on the blockchain and now owns a number of tokens.
Now assume the user wants to sell a certain amount of their tokens at a variable price that increases $1$ ether per day.
\figurename~\ref{fig:example} shows a possible implementation of an Ethereum smart contract using Solidity.
The smart contract allows a user to sell its tokens to an arbitrary user on the Ethereum blockchain.
The contract sells the tokens to the first buyer willing to pay $42$ ether, plus $1$ ether for each day since the start of the sale.
Moreover, the token sale should last no longer than $30$ days.
In this example, the smart contract acts as a simple mediator that automatically settles the trade between the user owning the tokens and the user willing to buy the tokens without both users requiring to know or trust each other. 
Smart contract based ICOs often follow a standard that is known as ERC-20~\cite{erc20}. 
This standard provides an interface that standardizes function names, parameters, and return values.
For example, the standard includes a function called \code{transferFrom}, which allows a user to transfer a limited amount of tokens to an arbitrary user on behalf of the owning user.
Another example is the function \code{allowance}, which returns the number of tokens that a user can spend on behalf of the owning user.
The smart contract in \figurename~\ref{fig:example} works as follows.
An arbitrary user can call the function \code{buy} to purchase the tokens for $42$ ether and a fee of $1$ ether for the number of days passed since the launch of the token sale.
The contract will automatically transfer the tokens by calling the function \code{transferFrom} on the ICO's contract.
After the purchase, the smart contract owner can call the function \code{withdraw} to retrieve the $42$ ether and the fee of the purchase.

The contract contains two vulnerabilities, one known as \emph{block dependency} and another known as \emph{leaking ether}.
The latter is enabled via a bug in the function \code{Tokensale} (see line 15 in \figurename~\ref{fig:example}).
Before Solidity version 0.4.22, the only way of defining a constructor was to create a function with the same name as the contract.
The function \code{Tokensale} is supposed to be the constructor of the contract \code{TokenSale}. 
Due to a typo, the names do not match, and the compiler does not consider the function as the contract's constructor.
As a result, the function \code{Tokensale} is considered a public function that any user on the blockchain can call.
This type of programming mistake has led to multiple attacks in the past \cite{atzei2017survey}.
The first vulnerability, namely block dependency, occurs when the transfer of ether depends on block information, such as the timestamp (see line 28 in \figurename~\ref{fig:example}). 
Malicious miners can alter the timestamp of blocks that they mine, especially if they can gain advantages. 
Although miners cannot set the timestamp smaller than the previous one, nor can they set the timestamp too far ahead in the future, developers should still refrain from writing contracts where the transfer of ether depends on block information.
The second vulnerability, namely leaking ether, occurs whenever a contract allows an arbitrary user to transfer ether, despite having never transferred ether to the contract before.
The following sequence of transactions triggers both vulnerabilities:

\begin{itemize}
    \item $t_0$: A non-malicious user calls the function \code{buy} with a value equals to 42 ether + fee; 
    \item $t_1$: An attacker calls the function \code{Tokensale};
    \item $t_2$: The same attacker calls the function \code{withdraw} after 30 days.
\end{itemize}

When running the above example using
\textsc{ILF} \cite{he2019ilf} (a state-of-the-art smart contract fuzzer), it is not capable of finding the two vulnerabilities even after 1 hour.
Inspecting the code coverage reveals that \textsc{ILF} achieves only $39$\%.
For comparison, \confuzzius{} achieves roughly 95\% code coverage and correctly identifies the two vulnerabilities in less than $10$ seconds.

\subsection{Input Generation}
\label{subsec:input_generation}

\begin{figure}
\centering
\includegraphics[width=1.0\columnwidth]{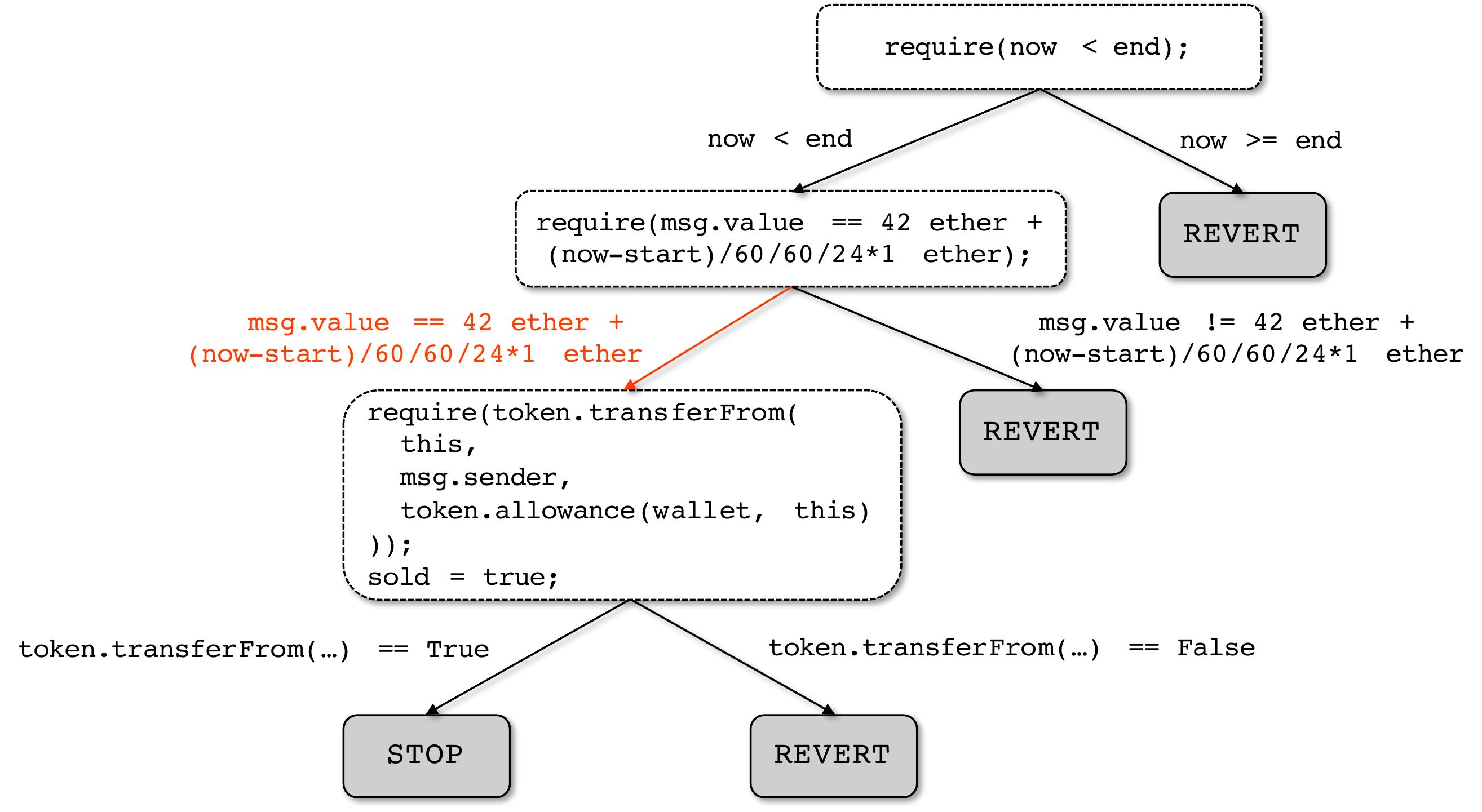}
\vspace*{-5mm}
\caption{Control-flow graph of the function \code{buy()}, where the complex condition is highlighted in red.}
\label{fig:cfg}
\end{figure}

Generating meaningful inputs is crucial for automated software testing.
Fuzzers generate inputs in order to execute not-yet-executed code. 
This generation can be completely random (black-box fuzzers) or driven by runtime information (grey-box fuzzers). 
In both cases, the primary approach is to mutate previous inputs to generate new inputs to test. 
Thus, finding the right heuristics is of fundamental importance to efficiently explore the target input space and, eventually, find latent bugs in the code.
However, real-world programs tend to contain conditions that are hard to trigger.
These complex conditions need to be addressed by fuzzers in order to execute as much code as possible. 
Line 21 in \figurename~\ref{fig:example} provides an example of a complex condition.
Function \code{buy} requires the transaction value to be equal to $42$ ether along with a variable fee that depends on the number of days that have past since the launch of the token sale.
\figurename~\ref{fig:cfg} illustrates the control-flow graph (CFG) of the function \code{buy} along with its branching conditions. 
The complex condition is highlighted in red in the CFG.
A fuzzer following a traditional random strategy will fail to get past this condition since it will generate the desired value only once every $2^{256}$ trials. 

Existing smart contract fuzzers such as \textsc{Harvey}~\cite{wustholz2019harvey} instrument the code and compute cost metrics for every branch to mutate the inputs. Our approach applies constraint solving to generate values for complex conditions on-demand. 
However, our fuzzer does not directly propagate these values, but instead, it stores them in so-called mutation pools.
Mutation pools manage a set of values that the fuzzer can use to get past complex conditions.
Every function has its own set of mutation pools, namely a mutation pool per function argument, transaction argument (e.g.~transaction value), and environmental argument (e.g.~block timestamp). 
Initially, all the pools are empty, and the fuzzer uses randomly generated values to feed the target functions. 
Once the fuzzer cannot discover new paths, it activates the constraint solver to generate new values.
We use symbolic taint analysis to create the expressions required by the constraint solver to generate new values.
We introduce taint in the form of a symbolic value whenever we come across an input during execution. 
This symbolic value is then propagated throughout the program execution, thereby forming step-by-step a symbolic expression that reflects the constraints on the particular input.
Solving these expressions will result in new values that will be added to the mutation pools. 
The fuzzer will then pick these values from the mutation pools and generate new inputs that execute new paths.
In \figurename~\ref{fig:example}, once \confuzzius{} realizes that the code coverage is not increasing, it activates the constraint solver, which outputs the value $42$ along with the current fee depending on the current block timestamp. It adds it to the mutation pool that manages the transaction value for the function \code{buy}. 
The value will be then picked up from the mutation pool by the fuzzer in the next round, and the execution of the transaction will evaluate the condition at line 21 to \code{True}, which results in getting past the missing branch and executing new lines of code.

\subsection{Stateful Exploration}
\label{subsec:state_generation}

Due to the transactional nature of blockchains, smart contract fuzzers must consider that each transaction may have a different output depending on the contract's current state, \ie all the previously executed transactions. 
Appropriately combining multiple transactions is necessary to generate states that trigger the execution of new branches. 
Ethereum smart contracts have, besides a volatile memory model, also a persistent memory model called \emph{storage}, which allows them to keep state across transactions.
For example, the global variables \code{end}, \code{wallet}, \code{token}, \code{owner}, and \code{sold} in \figurename~\ref{fig:example} are storage variables and their values might change across transactions.
Let us consider the two vulnerabilities mentioned earlier. 
An attacker will only be able to extract the funds via the function \code{withdraw}, if the two variables \code{owner} and \code{sold} contain the address of the attacker and \code{True}, respectively.
However, this is only possible if the functions \code{buy} and \code{Tokensale} are called before the function \code{withdraw}. 
Thus only a particular combination of the three functions will trigger the two vulnerabilities. 
Although this example may seem straightforward, automatically finding the right combination of function calls within contracts with many functions can become challenging as the number of possible combinations grows exponentially.
We base our solution on a simple observation: a transaction influences the output of a subsequent set of transactions if and only if it modifies a storage variable that one of the subsequent transactions will use. 
This property is a known data dependency called \emph{read-after-write} (RAW)~\cite{hennessy2011computer}.
In the first step, \confuzzius{} traces all the storage reads and writes performed by a transaction along with the storage locations.
Afterwards, \confuzzius{} combines transactions so that transaction $a$ is executed after transaction $b$, only if $a$ reads from the same storage location where $b$ writes to.
The fuzzer always executes the combination of transactions on a clean state of the contract.
Thus, a transaction sequence contains only transactions that change the state used by one of the subsequent transactions within the same sequence by construction. 
In the example of \figurename~\ref{fig:example}, \confuzzius{} will progressively learn that:
\begin{itemize}
\item \code{buy} reads variable \code{end} and writes to variable \code{sold};
\item \code{Tokensale} writes to variable \code{owner};
\item \code{withdraw} reads variable \code{owner} and variable \code{sold}.
\end{itemize}

\begin{figure}
\centering
\includegraphics[width=0.9\columnwidth]{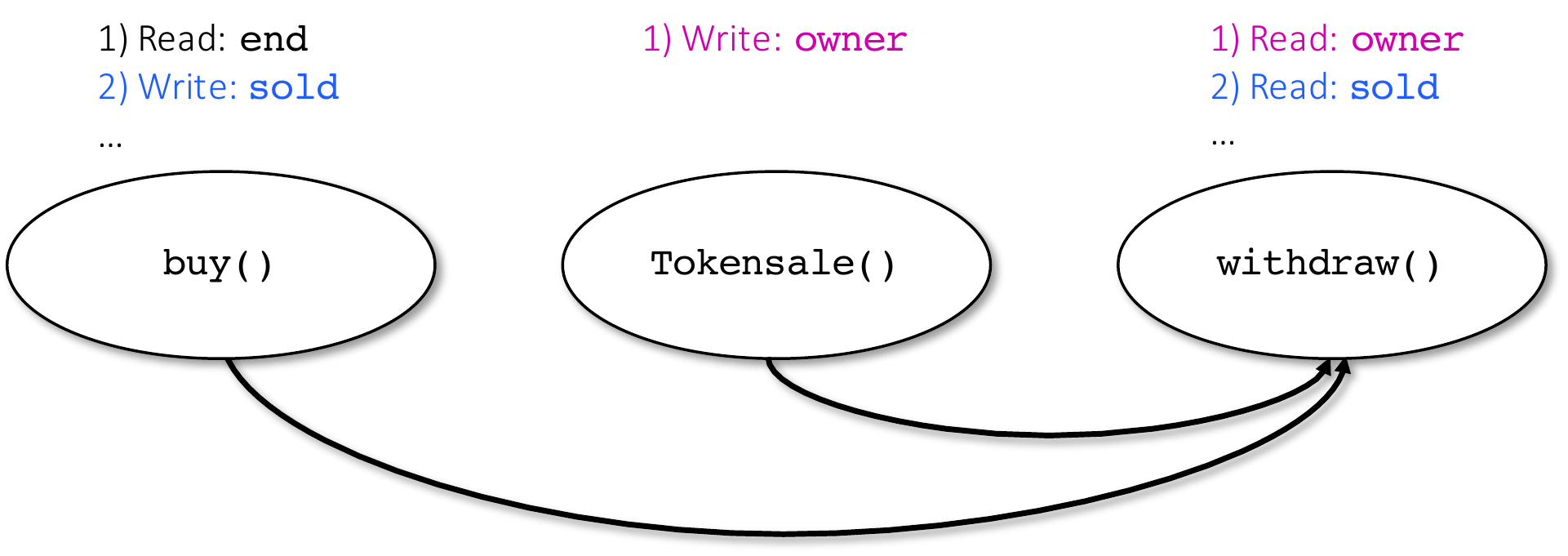}
\caption{A dependency graph illustrating the read-after-write (RAW) data dependencies contained in \figurename~\ref{fig:example}. A node represents a smart contract function and an edge indicates a RAW dependency between the two functions.}
\label{fig:raw_example}
\end{figure}

Using the information learned above and combining transactions based on RAW dependencies, \confuzzius{} will eventually create the following transaction sequence: 
$$\code{buy() } \rightarrow \code{ Tokensale() } \rightarrow \code{ withdraw()}$$
The directed graph in \figurename~\ref{fig:raw_example} presents the RAW dependencies to generate all the possible combinations. 
The graph shows that the functions \code{buy} and \code{Tokensale} must be executed before the function \code{withdraw}, but that the order between the two can be arbitrary. 

\subsection{Environmental Dependencies}
\label{sec:enironmental_challenge}

The execution of a smart contract does not only depend on the transaction arguments or the contract's current state. 
A smart contract's control-flow can also depend on input originating from the execution environment (\eg a block's timestamp).
Let us consider the contract in \figurename~\ref{fig:example}. 
Even though the function \code{withdraw} has no input argument,
the transfer of the balance is bound to some requirements. 
The requirement at line 28 is only satisfied if the transaction that triggered the function call is part of a block created 30 days after the contract's deployment.
Thus, the condition is bound to the mining mechanism of the Ethereum blockchain. 
While users submit transactions to the blockchain, miners aggregate them into blocks and distribute them to other nodes upon validation. 
When executing the transactions included in the block, the EVM accesses the block information contained therein. 
Block information includes the block hash, the miner's address, the block timestamp, the block number, the block difficulty, and the block gas limit. 
We solve this challenge by modeling this information as a fuzz-able input.
These inputs follow the same fuzzing procedure as transaction inputs.
We modified the EVM in order to be able to inject the fuzzed block information during the execution of the smart contract.
However, modeling block information as fuzz-able inputs is not enough. 
The EVM also permits to call other contracts deployed on the blockchain. Thus the control-flow of a smart contract may depend on the result of calling other contracts.
Consider line 29 in \figurename~\ref{fig:example}, where the state variable \code{sold} is required to be set to \code{True} in order for the attacker to be able to retrieve the funds.
The variable \code{sold} can only be set to \code{True} if the two contract calls at line 22 (\eg \texttt{token.allowance} and \texttt{token.transferFrom}) are successful.
We solve this challenge in a similar way by instrumenting calls to contracts and modeling return values as fuzz-able inputs. Our modified EVM then injects the fuzzed return values at runtime.

\section{Design and Implementation}
\label{sec:design_implementation}

\begin{figure*}
\centering
\includegraphics[width=0.9\textwidth]{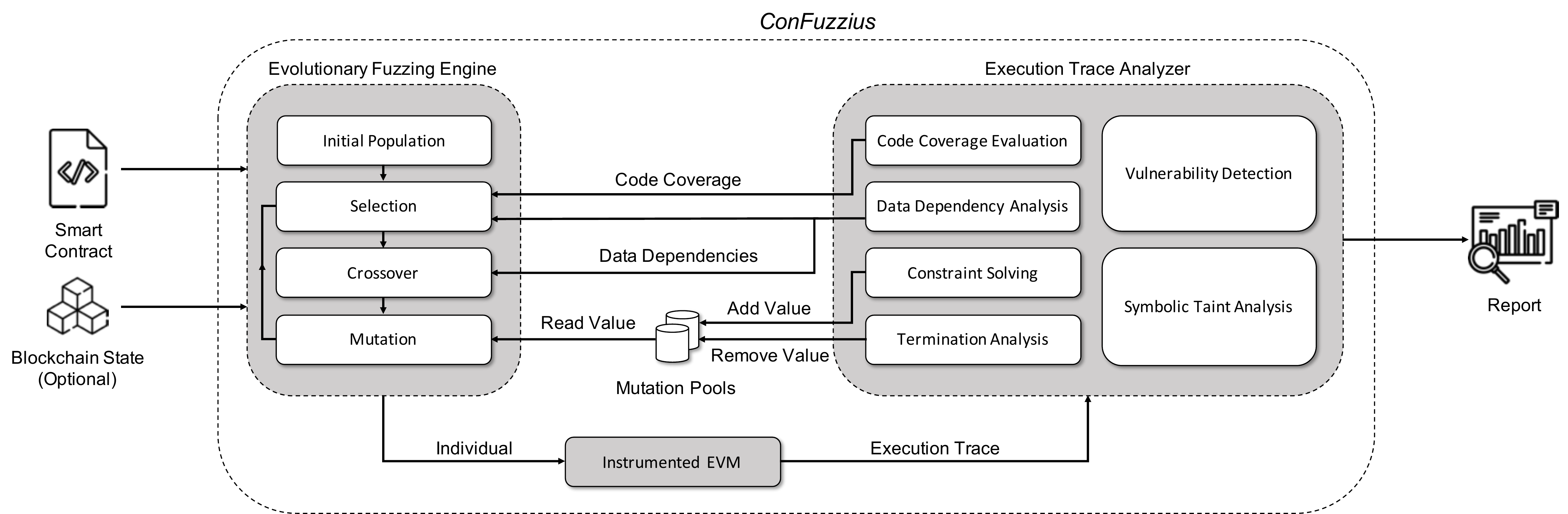}
\caption{Overview of \textsc{ConFuzzius}'s hybrid fuzzing architecture. The shadowed boxes represent the three main components and form together a feedback loop.}
\label{fig:architecture}
\end{figure*}

In this section, we provide details on the overall design and implementation of \confuzzius.

\subsection{Overview}

\confuzzius{}'s architecture consists of three main modules: the evolutionary fuzzing engine, the instrumented EVM, and the execution trace analyzer.
\figurename~\ref{fig:architecture} provides a high-level overview of \confuzzius{}'s architecture and depicts its individual components.
\confuzzius{} has been implemented in Python with roughly 6,000 lines of code\footnote{Source code is available at \url{https://github.com/christoftorres/ConFuzzius}.}.
\confuzzius{} takes as input the source code of a smart contract and a blockchain state. 
The latter is in the form of a list of transactions and is optional. The blockchain state is convenient for fuzzing already deployed smart contracts or contracts that need to be initialized with a specific state. 
\confuzzius{} begins by compiling the smart contract to obtain the Application Binary Interface (ABI) and the EVM runtime bytecode.
The evolutionary fuzzing engine then starts by generating individuals for the initial population, based on the smart contract's ABI. 
After that, the engine follows a standard genetic algorithm (i.e., selection, crossover, and mutation) and propagates the newly generated individuals to the instrumented EVM. 
The instrumented EVM then executes these individuals and forwards the resulting execution traces to the execution trace analyzer.
Next, the execution trace analyzer performs several analyses, \eg symbolic taint analysis, data dependency analysis, etc.
The execution trace analyzer is also responsible for triggering the constraint solver, running the vulnerability detectors, updating the mutation pools, and feeding information related to code coverage and data dependencies to the evolutionary fuzzing engine.
This process is repeated until at least one of the two termination conditions is met: a given number of generations has been generated, or a given amount of time has passed.
Finally, \confuzzius{} outputs a report containing information about the code coverage and the vulnerabilities that it detected.

\subsection{Evolutionary Fuzzing Engine}
\label{sec:evolutionary_fuzzing_engine}

In the following, we provide details on the encoding, initialization, fitness evaluation, selection, combination, and mutation of individuals.

\begin{figure*}
\centering
\includegraphics[width=0.85\textwidth]{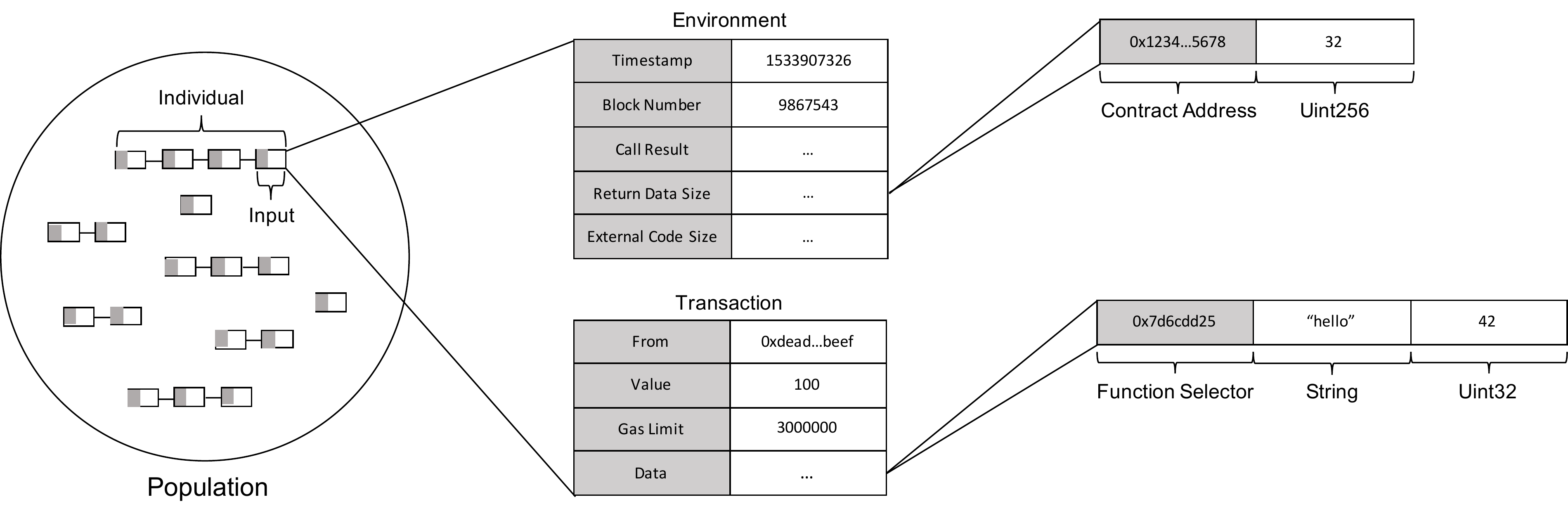}
\caption{Encoding of our population and its individuals. The shadowed boxes depict immutable values, whereas the non-shadowed boxes depict mutable ones.}
\label{fig:individual_encoding}
\end{figure*}

\noindent
\newline
\textbf{Encoding Individuals.} One of the most important decisions to make while implementing an evolutionary fuzzer is deciding on the representation of individuals.
Improper encoding of individuals can lead to poor performance~\cite{liepins1990representational}.
\figurename~\ref{fig:individual_encoding} illustrates our encoding of individuals.
Vulnerabilities are usually triggered either by sending a single transaction or a sequence of transactions to a smart contract.
However, transactions alone are not enough to trigger vulnerabilities (see Section~\ref{sec:enironmental_challenge}).
Specific vulnerabilities depend on the execution environment to be in a specific state.
Thus, our encoding represents an individual as a sequence of inputs.
Every input consists of an environment and a transaction. 
Both are encoded as key-value mappings.
The environment includes block information such as the current timestamp and block number, but it also includes call return values, data sizes, and external code sizes.
The latter three are encoded as an array of mappings, where a contract address maps to a mutable value (\eg a call result or a size).
The transaction includes the address of the sending account (\emph{from}), the transaction amount (\emph{value}), the maximum amount of gas for the contract to execute (\emph{gas limit}), and the input data for the contract to execute (\emph{data}). 
The input data is represented as an array of values where the first element is always the function selector, and the remaining elements represent the function arguments. 
The function selector is computed using the ABI and extracting the first four bytes of the Keccak (SHA-3) hash of the function signature.
As an example, the function \code{test(string a, uint b)}, has the string \code{test(string,uint)} as its function signature, which after hashing and extracting the first four bytes, results in \code{0x7d6cdd25} being its function selector.

\noindent
\newline
\textbf{Initial Population.}
The population is initialized with $N$ individuals, each of which initially contains only a single input (i.e.~a single transaction).
The function selector to be included in the transaction is selected in a round-robin fashion.
Function arguments are generated based on their type, which we obtain through the ABI.
Depending on the type and size (\ie fixed or non-fixed) of the argument, we apply different strategies to generate valid arguments for each function.
For example, if the argument type is a fixed size \code{uint32}, then we randomly choose a value either from the valid input domain (\eg between $0$ and $2^{32}-1$) or from a set of inputs that trigger edge cases of the valid input domain (\eg $0$, $1$, $4294967295$). 
The population is reinitialized whenever there has been no increase in code coverage for the past $k$ generations. This soft reboot introduces back diversity into the population and procrastinates premature convergence when the population has become homogeneous.

\noindent
\newline
\textbf{Fitness Evaluation.} 
The fitness evaluation of individuals plays a crucial role in evolutionary fuzzing.
The computation of the fitness function is done repeatedly and must be therefore sufficiently fast. 
A slow computation can adversely make the fuzzing exceptionally slow.
The fitness function is supposed to represent the landscape of the problem.
In general, evolutionary fuzzers aim to achieve complete coverage of the code.
While obtaining full code coverage does not necessarily mean that all vulnerabilities will be found, it is undoubtedly true that no vulnerabilities will be found in code that has not been explored.
Our fitness function is based on branch coverage (a form of code coverage) and data dependencies. 
We define our fitness function for an individual $i$ as follows:

\begin{equation} \label{equ:fitness_function}
fit(i) = fit_{branch}(i) + fit_{RAW}(i)
\end{equation}

\noindent
The fitness $fit_{branch}$ is computed by counting the number of branches that remain unexplored by the individual.
We keep track of all the branches that have been executed so far by all the individuals.
Then, we iterate through the execution trace of the individual and
analyze every conditional jump instruction (\ie \code{JUMPI} instruction).
A conditional jump always has two destinations, one for the \code{True} branch and one for the \code{False} branch.
We obtain the jump destination of the \code{True} branch by extracting it from the stack and the jump destination of the \code{False} branch by incrementing the program counter by one.
We increase the individual's fitness value $fit_{branch}$ by one for every jump destination that is not in our list of executed branches. 
This approach aims to prioritize individuals that require more exploration since these individuals will allow us to explore new parts of the contract.
However, this metric alone is not enough. We are also interested in preserving individuals that allow us to create useful sequences of transactions (\eg sequences with RAW dependencies), even though these individuals might have been already explored extensively.
Therefore, we compute the fitness $fit_{RAW}$, which takes this into account by using the data dependencies detected by our execution trace analyzer.
We start with a $fit_{RAW}$ of zero and increment $fit_{RAW}$ by one for every write to storage that the individual has performed during execution.
The final fitness value of an individual is then defined by the the sum of the two fitness values $fit_{branch}$ and $fit_{RAW}$. The combination of these two values allows us to drive the genetic algorithm to explore unexplored code while preserving individuals that are useful to explore deeper states of the smart contract.

\noindent
\newline
\textbf{Selection.} 
The process of choosing two individuals for the crossover step is called selection.
Literature proposes several selection operators~\cite{sivaraj2011review}.
We choose linear ranking selection as our selection operator since it considers the population as a whole during selection and not just a subset as it is, for example, the case in tournament selection.
In linear ranking selection, individuals with a high fitness value are ranked higher than those with a low fitness value. 
In other words, an individual is selected with a probability that is linearly proportional to the individual's rank in the population.
Hence, the worst individual has a rank of 1, the second-worst a rank of 2, the best-performing individual has a rank of $ N $, where $ N $ is the size of the population. 
All individuals have a chance of being selected, although the higher-ranked individuals will be slightly preferred. 
However, traditional linear ranking selection does not consider data dependencies between individuals.
Therefore, we propose a modified version of the linear ranking selection strategy, where the first individual is selected based on linear ranking selection, however, the second individual is selected based on having a RAW dependency with the first individual following a round-robin fashion.
In case there is no individual to have a RAW dependency with the first individual, we fallback to traditional linear ranking selection to select the second individual.

\noindent
\textbf{Crossover.} 
The crossover operator creates two new individuals by recombining the input sequences of two existing individuals.
Instead of randomly combining two individuals, we combine an individual after another only if the first performs a write to a storage location from which the second performs a read (RAW dependency).
There are only two possible combinations in our case: individual $a$ appended to individual $b$, or vice versa, individual $b$ appended to individual $a$.
If a combination yields a RAW dependency, then we combine both individuals by first selecting the individual whose input sequence performs the write and then append the individual whose input sequence performs the read.
As opposed to traditional crossover, we are concatenating individuals rather than splitting them apart and swapping their input sequences.
This preserves the RAW dependencies within the individuals themselves and creates individuals with new RAW dependencies.
If there is no RAW dependency between two individuals, then we simply return one of the two individuals unmodified.
However, it should be noted that individuals are not always combined even though they might have a RAW dependency. Individuals are combined based on a given crossover probability $p_c$.
Moreover, to prevent individuals from growing indefinitely large, we check before combining if the sum of their lengths exceeds the maximum size of $l$, and only combine them if their lengths are lower or equal to $l$. 
On the one hand, a small $l$ will produce shorter combinations of individuals, which will result in shorter execution times but also in finding less bugs. 
On the other hand, a larger $l$ will result in longer combinations of individuals, which will result in finding more bugs that lay deeper in the execution, but also in longer execution times. 
Therefore, a trade-off between completeness and performance must be taken when selecting an appropriate value for $l$.

\noindent
\newline
\textbf{Mutation.}
The mutation operator randomly modifies parts of a single individual in order to create a new individual. 
It introduces diversity in the population.
Our mutation operator works by iterating through the sequence of inputs of an individual and mutating every environmental and transactional value based on a shared mutation probability $p_m$.
A value can be mutated in two ways: replacing the original value with a random one or replacing the original value with a value from a \textit{mutation pool}.
Mutation pools act as a form of short-term memory. 
They allow the fuzzer to reuse values that have been previously observed or learned during past executions.
There are in total nine different mutation pools, one per transactional and environmental value type. 
Hence, our fuzzer has a mutation pool for senders, amounts, gas limits, function arguments, timestamps, block numbers, call results, call data sizes, and external code sizes.
All mutation pools are implemented to map a function selector to a circular buffer, except for the mutation pool on function arguments.
The implementation is similar, except that we do not directly map the function selector to a circular buffer but to another mapping that maps to an argument index and then to a circular buffer. 
Thus, the pool for function arguments first maps to a function selector, then to an argument index, and then to a circular buffer. 
This is because functions can have more than just one argument, and we want to keep track of interesting values for every argument separately.
Circular buffers help us ensure that the values contained therein are rotated in a round-robin like fashion. Old values are overwritten by newer ones (\ie mimicking short-term memory). 
Our buffers can hold up to $10$ values by default.
All mutation pools are initially empty, except for the mutation pool tracking transaction amounts, which gets initialized with the values 0 and 1.
When mutating a transactional or environmental value, we first check if the associated mutation pool is empty.
If the pool is empty, we inject a randomly generated value based on the type of information extracted from the ABI.
Otherwise, we inject the current value at the head of the circular buffer and rotate it. 

\subsection{Instrumented EVM}

The EVM is responsible for executing the transactions generated by the individuals on the runtime bytecode of the contract that is under test. 
Its efficiency has a significant impact on the overall performance of the fuzzer.
Hence, the EVM must achieve a high processing rate of transactions.
Every official Ethereum client implementation allows users to deploy smart contracts locally and send transactions to them. 
However, all of these clients require transactions to be encoded using the Recursive Length Prefix (RLP) format in order to be mined.
We realized that the actual EVM execution time is negligible compared to the effort of encoding and decoding a transaction to and from the RLP format.
Therefore, we decided to reuse an official Python implementation of the EVM~\cite{pyEVM} and incorporate it within our fuzzer.
This removes the burden of mining blocks as well as encoding transactions and thus significantly speeds up the execution.
Moreover, we slightly modified the EVM in order to be able to retrieve the execution trace of a transaction. 
An execution trace consists of an array, where every element contains the name of the executed instruction, the program counter, the execution stack, the call-stack depth, and a flag stating if an internal error has occurred during execution.
The EVM itself is by default stateless and uses the blockchain to preserve states. 
However, since we are not interested in the internal persistence mechanism of the Ethereum blockchain, we decided to implement a simple storage emulator that is used by our EVM to persist the state changes that are performed during execution.
All state changes are kept in memory to further improve the speed of execution.
Besides persisting the state of smart contract executions, the storage emulator also allows us to inject custom environmental information such as the block timestamp or modify the result of a call.
Finally, the storage emulator also enables us to create snapshots of the current state of the EVM.
This allows us to quickly reset the state of the EVM to an initial state without having to redeploy the smart contract every time from scratch when executing the transactions of an individual.

\subsection{Execution Trace Analyzer}
\label{sec:execution_trace_analyzer}

The execution trace analyzer receives from the instrumented EVM the execution trace and then performs a number of analyses, such as code coverage evaluation, data dependency analysis, symbolic taint analysis, vulnerability detection, constraint solving, and termination analysis.
Moreover, the execution trace analyzer also manages the values stored within the mutation pools and is responsible for providing code coverage information and data dependencies to the evolutionary fuzzing engine.
Finally, it is also in charge of generating a report containing statistics about code coverage and vulnerabilities.

\noindent
\newline
\textbf{Code Coverage Evaluation.} 
Code coverage is necessary for computing an individuals' fitness and detecting when the evolutionary fuzzing engine should activate constraint solving or reset the population. 
The code coverage is computed by counting the number of unique program counter values within the execution trace.
 
\begin{table}
\centering
\caption{Storage Layout of State Variables in Solidity}
\begin{adjustbox}{width=\columnwidth,center}
\begin{tabular}{| l l l l |}
\hline
    \textbf{Variable Type} & 
    \textbf{Declaration} & 
    \textbf{Access} & 
    \textbf{Storage Location} \\
\hline \hline
Primitive & \code{T v} & \code{v} & \code{s}(\code{v}) \\
Struct & \code{struct} \code{v} \{ \code{T a} \} & \code{v}.\code{a} & \code{s}(\code{v}) + \code{s}(\code{a}) \\
Fixed Array & \code{T[10] v} & \code{v[n]} & \code{s}(\code{v}) + \code{n} $\cdot$ $\vert$\code{T}$\vert$ \\
Dynamic Array & \code{T[] v} & \code{v[n]} & \code{h}(\code{s}(\code{v})) + \code{n} $\cdot$ $\vert$\code{T}$\vert$ \\
	& & \code{v}.\code{length} & \code{s}(\code{v}) \\
Mapping & \code{mapping(T$_1$ => T$_2$)} \code{v} & \code{v[k]} & \code{h}(\code{k} $\vert\vert$ \code{s}(\code{v})) \\
\hline
\end{tabular}
\end{adjustbox}
\label{tbl:storage}
\end{table}
 
\noindent
\newline
\textbf{Data Dependency Analysis.} 
Fitness evaluation, selection, and crossover require information about data dependencies. 
The data dependency analysis tracks all the state variables, read from and written to, throughout the execution of the fuzzer.
In contrast to existing approaches~\cite{zhang2019mpro}, which extract data dependencies via static analysis, our fuzzer retrieves the access patterns to state variables at runtime (\ie dynamically) by iterating through the execution trace and scouting for \code{SLOAD} and \code{SSTORE} instructions.
The advantage of static analysis is that it is fast compared to dynamic analysis. 
However, the disadvantage is that it requires source code to precisely track data flows across variables.
While data flows could be extracted from bytecode through static analysis, it would only work for simple variable types such as primitives and not for complex types such as mappings or arrays. 
Dynamic analysis on the other hand, allows us to track variables with complex types, even without source code. The disadvantage is the additional runtime overhead and implementational effort. 
The instruction \code{SLOAD} denotes a read from storage, whereas an \code{SSTORE} instruction denotes a write to storage.
Table~\ref{tbl:storage} depicts how Solidity computes the storage location for different types of state variables~\cite{storagelayout}.
The function \code{s}() determines the so-called \emph{storage slot} of a particular variable \code{v}, whereas the function \code{h}() computes a Keccak-256 hash.
Statically-sized variables such as primitives, structs, and fixed-size arrays, are laid out contiguously in storage starting from position 0.
Solidity uses a Keccak-256 hash computation to define the stored data's starting position due to the unpredictable size of dynamically-sized arrays and mappings.
However, we are not interested in identifying individual storage locations but rather access to a particular variable.
Therefore, our goal is to extract the storage slot \code{s}(\code{v}) for a variable \code{v}, instead of the storage location. As an example, assume we have a variable called \code{balances} of type mapping, which maps an \code{address} to a \code{uint}, and we have two addresses \code{a} and \code{b}. The storage location for \code{balances[a]} and \code{balances[b]} will be different, since the computation of the storage locations will be \code{h}(\code{a} $\vert\vert$ \code{s}(\code{balances}) and \code{h}(\code{b} $\vert\vert$ \code{s}(\code{balances})), respectively, where $\vert\vert$ means concatenation. However, these two storage locations share the same storage slot 
\code{s}(\code{balances}), which enables us to link both storage locations together to the same variable \code{balances}.
Extracting the storage slot for statically-sized variables is straightforward as it is equivalent to the storage location.
It can be achieved by merely popping the first element from the stack for both instructions, \code{SLOAD} and \code{SSTORE}.
Extracting storage slots for mappings and dynamic arrays is more challenging. 
As it is not possible to invert the result of a hash, we must keep track of the Keccak-256 hash computations by mapping the result of a \code{SHA3} instruction to the memory contents that were involved in the computation.
Note that the \code{SHA3} instruction computes the Keccak-256 hash from a memory slice determined via two arguments on the stack, namely the memory offset and the memory size. 
Thus, for mappings, all we need to do is to obtain the mapped memory contents.
To deal with concatenation, we only extract the last 32 bytes of the memory contents as these represent the storage slot.
Finally, for dynamic arrays, we must keep track of the arithmetic addition of Keccak-256 hashes. 
The storage slot is then determined the same way as for mappings, except that there is no concatenation.

\noindent
\newline
\textbf{Symbolic Taint Analysis.}
The symbolic taint analysis produces symbolic constraints that are later used by other parts, such as constraint solving and vulnerability detection. 
We introduce taint in the form of symbolic values and track their flow across instructions.
We leverage light dynamic taint analysis by injecting taint only for instructions that can be fuzzed, \eg \code{CALLDATALOAD}, \code{CALLVALUE}, or \code{TIMESTAMP}.
Taint is propagated across stack, memory, and storage of the execution trace.
The propagation of taint across storage allows us to do inter-transactional taint analysis.
We implemented the stack using an array structure that follows LIFO logic. 
We used a Python dictionary that maps memory and storage addresses to values to represent memory and storage. 
Since the EVM is a stack-based register-less virtual machine, the operands of instructions are always passed via the stack. 
Therefore, our taint propagation method identifies each EVM bytecode instruction's operands and propagates the taint according to each instruction's semantics as defined in the yellow paper \cite{wood2014ethereum}. 
The taint propagation logic follows an over-tainting policy, which tags the instruction's output as tainted if at least one of the instruction's inputs are tainted.

\noindent
\newline
\textbf{Constraint Solving.}
There are situations where the evolutionary fuzzing engine converges prematurely because it cannot advance past a complex conditional statement.
The constraint solver's role is then to generate a valid input that allows the evolutionary fuzzing engine to get past the complex condition. 
The symbolic taint analysis tries to build a logical formula that describes the complex execution path, thereby reducing the problem of reasoning about the execution to the domain of logic.
These logical formulas are often called path constraints. 
We implemented our own lightweight symbolic execution engine, that only executes instructions related to arithmetic operations (\eg\code{ADD}, \code{MOD}, \code{EXP}), comparison logic (\eg\code{LT}, \code{EQ}) and bitwise logic (\eg\code{AND}, \code{NOT}). 
The engine consists of an interpreter loop that gets instructions from the execution trace and symbolically executes them. 
The loop continues until all the instructions contained in the execution trace have been executed.
We obtain the formulas only for conditional statements with open branches, \ie never executed branches.  
Each formula contains the path constraints to reach the conditional statement.
We negate the last constraint, substitute the symbolic variables in the rest of the logical formula with concrete values that have been used as inputs to trigger the execution trace, and use the Z3 SMT solver~\cite{de2008z3} to produce inputs to reach the open branch.
Concretization helps us reduce the complexity of the formula and therefore avoid the path explosion problem.
We then add the produced inputs to the mutation pools.
Eventually, in one of the following generations, the mutation operator will pick up the solution, and our evolutionary fuzzing engine will now be able to get past the complex condition.
We also add the previously used inputs to the mutation pools, allowing the fuzzer to execute both branches.

\noindent
\newline
\textbf{Termination Analysis.}
The execution traces may contain valuable feedback on the validity of the inputs.
Our fuzzer uses the execution traces to obtain feedback and learn whether an input is meaningful or not. 
The termination analysis inspects the execution traces for opcodes that indicate either correct or incorrect termination of execution.
Invalid inputs will result in the execution trace terminating with a \code{REVERT}, \code{INVALID}, or \code{ASSERTFAIL} instruction, whereas valid inputs will result in the execution terminating with either a \code{SELFDESTRUCT}, \code{SUICIDE}, \code{STOP}, or \code{RETURN} instruction.
Once we detect an incorrect termination, we analyze the last path constraint before the termination and retrieve the input values responsible for the incorrect termination. We then remove these values from the mutation pools.
A transaction that results in an incorrect termination reverts all state changes made during execution and is therefore not relevant for creating meaningful RAW dependencies. This helps the fuzzer focus on more relevant parts of the code.

\noindent
\newline
\textbf{Vulnerability Detection.} 
We detect vulnerabilities by analyzing the execution traces and the information returned by the symbolic taint analysis and the data dependency analysis.
We define a detector per vulnerability. 
We implemented detectors for the following types of vulnerabilities: assertion failure (AF), integer overflow (IO), reentrancy (RE), transaction order dependency (TD), block dependency (BD), unhandled exception (UE), unsafe delegate call (UD), leaking ether (LE), locking ether (LO), and unprotected self-destruct (US).
A detailed explanation of each vulnerability and its detector is described in the Appendix A.
More detectors can be easily added to extend the detection capabilities of our fuzzer.
\begin{table*}
\centering
\caption{Security tools evaluated in this work. Tools marked with $\CIRCLE$ support the detection of the vulnerability, while tools marked with $\Circle$ do not support the detection of the vulnerability.%
}
\begin{tabular}{| l | c | c | c | c c c c c c c c c c |}
\hline
    \multirow{2}{*}{\textbf{Toolname}} & 
    \multirow{2}{*}{\textbf{Type}} & 
    \multirow{2}{*}{\textbf{Requires Source Code}} & 
    \multirow{2}{*}{\textbf{Requires ABI}} & \multicolumn{10}{c|}{\textbf{Vulnerability Detectors}} \\
    & & & &
    \textbf{AF} &
    \textbf{IO} &
    \textbf{RE} & 
    \textbf{TD} &
    \textbf{BD} &
    \textbf{UE} &
    \textbf{UD} &
    \textbf{LE} &
    \textbf{LO} &
    \textbf{US} \\
\hline \hline
\textsc{Oyente}~\cite{Luu2016}      & Symbolic & \xmark & \xmark & $\CIRCLE$ & $\CIRCLE$ & $\CIRCLE$ & $\CIRCLE$ & $\CIRCLE$ & $\Circle$ & $\Circle$ & $\Circle$ & $\Circle$ & $\CIRCLE$ \\
\textsc{Mythril}\cite{mueller2018}      & Symbolic & \xmark & \xmark & $\CIRCLE$ & $\CIRCLE$ & $\CIRCLE$ & $\CIRCLE$ & $\CIRCLE$ & $\CIRCLE$ & $\CIRCLE$ & $\CIRCLE$ & $\Circle$ & $\CIRCLE$ \\
\textsc{M-Pro}\cite{mueller2018}      & Symbolic & \cmark & \xmark & $\CIRCLE$ & $\CIRCLE$ & $\CIRCLE$ & $\CIRCLE$ & $\CIRCLE$ & $\CIRCLE$ & $\CIRCLE$ & $\CIRCLE$ & $\Circle$ & $\CIRCLE$ \\
\textsc{ILF}~\cite{he2019ilf}     & Fuzzer     & \cmark & \cmark & $\Circle$ & $\Circle$ & $\Circle$    & $\Circle$ & $\CIRCLE$ & $\CIRCLE$ & $\CIRCLE$ & $\CIRCLE$ & $\CIRCLE$ & $\CIRCLE$ \\
\textsc{sFuzz}\cite{nguyen2020sfuzz} & Fuzzer & \cmark & \cmark & $\Circle$ & $\CIRCLE$ & $\CIRCLE$ & $\Circle$ & $\CIRCLE$ & $\CIRCLE$ & $\CIRCLE$ & $\Circle$ & $\CIRCLE$ & $\Circle$ \\
\textbf{\textsc{ConFuzzius}} & \textbf{Hybrid} & \xmark & \cmark & $\CIRCLE$ & $\CIRCLE$ & $\CIRCLE$ & $\CIRCLE$ & $\CIRCLE$ & $\CIRCLE$ & $\CIRCLE$ & $\CIRCLE$ & $\CIRCLE$ & $\CIRCLE$ \\
\hline
\end{tabular}
\label{tbl:tools}
\end{table*}

\section{Evaluation
\label{sec:evaluation}}

In this section, we evaluate the effectiveness and performance of \confuzzius{} by answering the following three research questions:
\begin{description}
    \item[\textbf{RQ1}] Does \confuzzius{} achieve higher code coverage than current  state-of-the-art symbolic execution and fuzzing tools for smart contracts?
    \item[\textbf{RQ2}] Does \confuzzius{} discover more vulnerabilities than current state-of-the-art symbolic execution and fuzzing tools for smart contracts?
    \item[\textbf{RQ3}] How relevant are \confuzzius{}'s individual components in terms of code coverage and vulnerability detection?
\end{description}

\begin{table}
\addtolength{\tabcolsep}{-3.7pt}
\centering
\caption{
Statistics of our clusters and overall dataset.
}
\begin{tabular}{| c | r r r | r r r | r r r |}
\hline
    \multirow{2}{*}{\textbf{Dataset}} & 
    \multicolumn{3}{c|}{\textbf{LoSC}} & 
    \multicolumn{3}{c|}{\textbf{Functions}} &
    \multicolumn{3}{c|}{\textbf{Instructions}} \\
    & \textbf{Min} & \textbf{Max} & \textbf{Mean} & \textbf{Min} & \textbf{Max} & \textbf{Mean} & \textbf{Min} & \textbf{Max} & \textbf{Mean} \\
\hline \hline
Small & 1 & 3,584 & 119 & 1 & 81 & 14 & 1 & 3,632 & 1,763 \\
Large & 44 & 3,148 & 429 & 1 & 190 & 32 & 3,633 & 16,889 & 5,495 \\
\textbf{Overall} & \textbf{1} & \textbf{3,584} & \textbf{168} & \textbf{1} & \textbf{190} & \textbf{17} & \textbf{1} & \textbf{16,889} & \textbf{2,353} \\
\hline
\end{tabular}
\label{tbl:dataset}
\end{table}

\noindent
\newline
\textbf{Datasets.}
We run our experiments using two different datasets. The purpose of the first dataset is to measure code coverage, whereas the second dataset aims to measure the detection of vulnerabilities\footnote{Both datasets are available at \url{https://github.com/christoftorres/ConFuzzius}.}.
The first dataset was obtained by crawling Etherscan's list of verified smart contracts~\cite{verifiedContracts}.
These are real-world smart contracts where the source code is publicly available and verified to match the bytecode deployed on the Ethereum blockchain.
We filtered out contracts that failed to compile using Solidity version 0.4.26, resulting in a dataset of 21,147 contracts.
Moreover, we split our dataset into different clusters based on each contract's number of EVM bytecode instructions. The idea is to examine code coverage on smart contracts with different sizes.
We used the standard k-means clustering algorithm to create the clusters.
The number of clusters has been determined using the Elbow and the Silhouette method. 
Both methods yield $2$ to be the optimal value for $k$.
Table~\ref{tbl:dataset} lists the number of lines of source code (LoSC), number of public functions, and the number of EVM bytecode instructions for each cluster and the overall dataset.
The first cluster represents small contracts (\code{$\leq$} 3,632 instructions) and contains 17,803 contracts, whereas the second cluster represents large contracts (\code{$>$} 3,632 instructions) and contains 3,344 contracts.
The second dataset is based on Durieux et al.'s curated dataset
\cite{durieux2020empirical}, a collection of annotated smart contracts, which the authors used to evaluate the effectiveness of smart contract analysis tools. 
Unfortunately, their curated dataset is missing certain types of vulnerabilities (e.g. assertion failures).
We decided to reuse their dataset and extend it with annotated vulnerabilities from the Smart Contract Weakness Classification (SWC) registry~\cite{swcRegistry}.
We added contracts related to assertion failures, unsafe delegatecalls, leaking ether, locking ether, and unprotected selfdestructs. The extended dataset consists of 128 contracts with 148 annotated vulnerabilities.

\noindent
\newline
\textbf{Baselines.}
We compare \confuzzius{} to the tools listed in Table~\ref{tbl:tools}.
We limit our comparison to symbolic execution tools and fuzzers as we want to know if our hybrid approach performs better than these methods on their own.
We chose \textsc{Oyente}~\cite{Luu2016}
because of its popularity among the community and continuous development.
We chose \textsc{Mythril} since a recent study on smart contract analysis tools \cite{durieux2020empirical} revealed that it performs better than a variety of existing tools (e.g. Manticore~\cite{manticore}, SmartCheck~\cite{tikhomirov2018smartcheck}, Securify~\cite{tsankov2018securify}, Maian~\cite{nikolic2018finding}, etc.).
We chose \textsc{M-Pro} because it employs a similar transaction sequence combining strategy than ours, with the difference being that ours is dynamic, and theirs is static.
We chose \textsc{ILF}~\cite{he2019ilf} since it has proven to outperform existing smart contract fuzzers (i.e. \textsc{ContractFuzzer} \cite{jiang2018contractfuzzer} and \textsc{Echidna} \cite{echidna2020}).
We chose \textsc{sFuzz} because it is a recent work that has not been compared yet by previous works and because it is based on the popular fuzzing tool AFL.
Table~\ref{tbl:tools} compares the different types of vulnerabilities detected by each tool. 
We see that none of the tools are currently able to detect all of the ten vulnerabilities that are currently detectable by \confuzzius{}. We also see that \confuzzius{} does not requires source code as compared to the other fuzzers. 

\begin{figure}[t]
\centering
\includegraphics[width=0.49\textwidth]{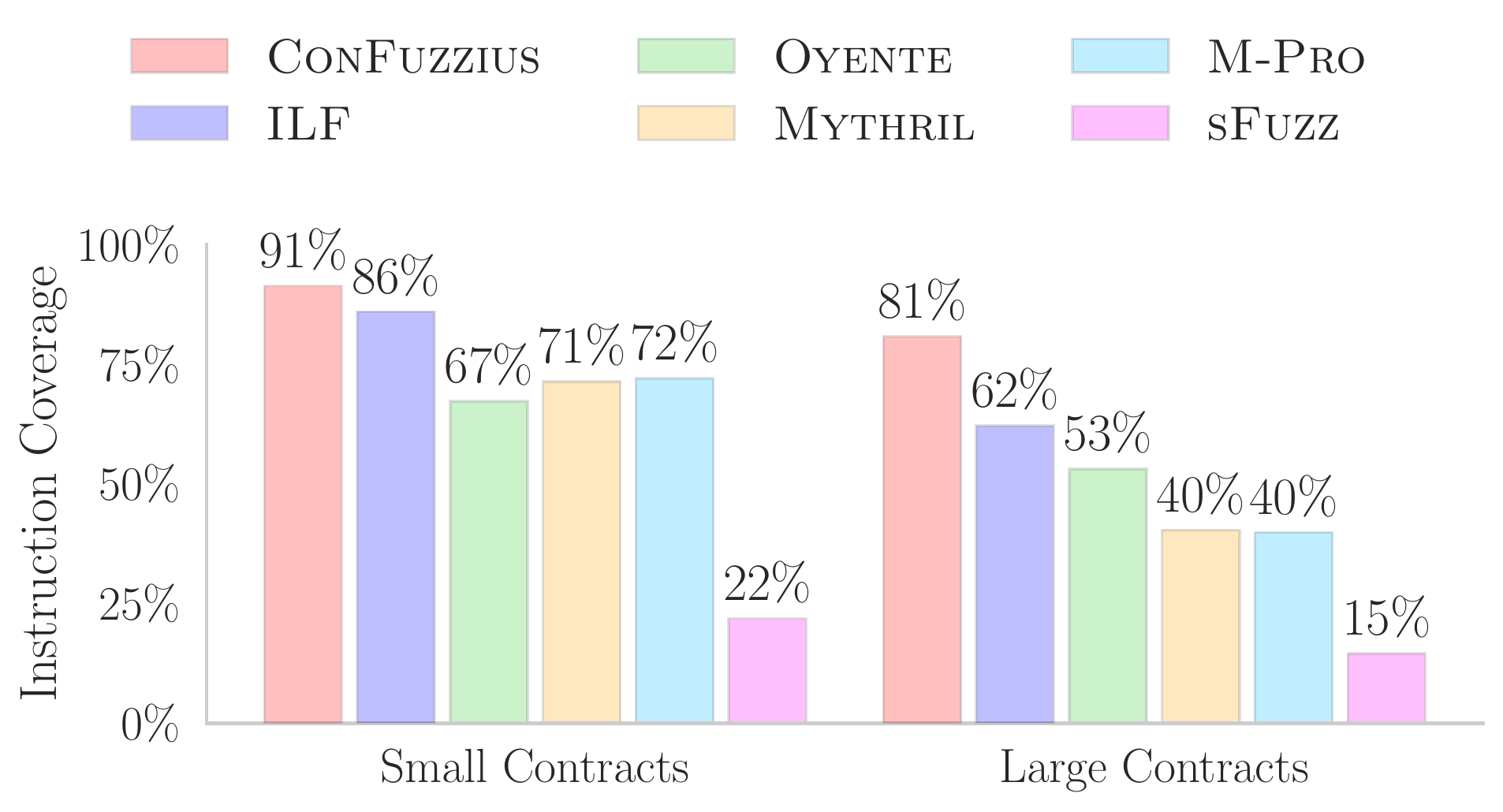}
\vspace*{-5mm}
\caption{Overall instruction coverage of \confuzzius{} and other tools.}
\label{fig:code_coverage}
\end{figure}

\noindent
\newline
\textbf{Experimental Setup.}
We followed the guidelines by Klees et al.~\cite{klees2018evaluating} on evaluating fuzz testing. 
For each experiment, we performed 10 runs, each with independent seeds.
For the first dataset we run experiments on the small contracts for 10 minutes each, whereas on the large contracts for 1 hour each.
Preliminary tests showed that most tools did not yield more coverage past these times.
Moreover, before every run, we initialized \confuzzius{}'s and \textsc{ILF}'s blockchain state with the same values that were used to deploy the contracts on the Ethereum mainnet.
For the second dataset, we run the experiments for each contract for 10 minutes.
We run our experiments on a cluster of 10 nodes, each with 128 GB of memory. 
Every node runs CentOS release 7.6.1810 and has 2 Intel\textsuperscript{\textregistered} Gold 6132 CPUs with 14 cores, each clocked at 2.60 GHz.
We run \confuzzius{} with a variable population size, that was computed as two times the number of functions contained in the ABI of the contract under test. 
We set the crossover probability and the probability of mutation to $0.9$ and $0.1$, respectively. 
The population is reinitialized whenever the code coverage does not increase for $k = 10$ generations.
We set the maximum length for individuals to $l = 5$. 
Finally, we used Z3 version 4.8.5 as our constraint solver with a timeout of 100 milliseconds per Z3 request.

\subsection{Code Coverage}

\begin{figure}
\centering
\vspace*{-2mm}
\includegraphics[width=0.49\textwidth]{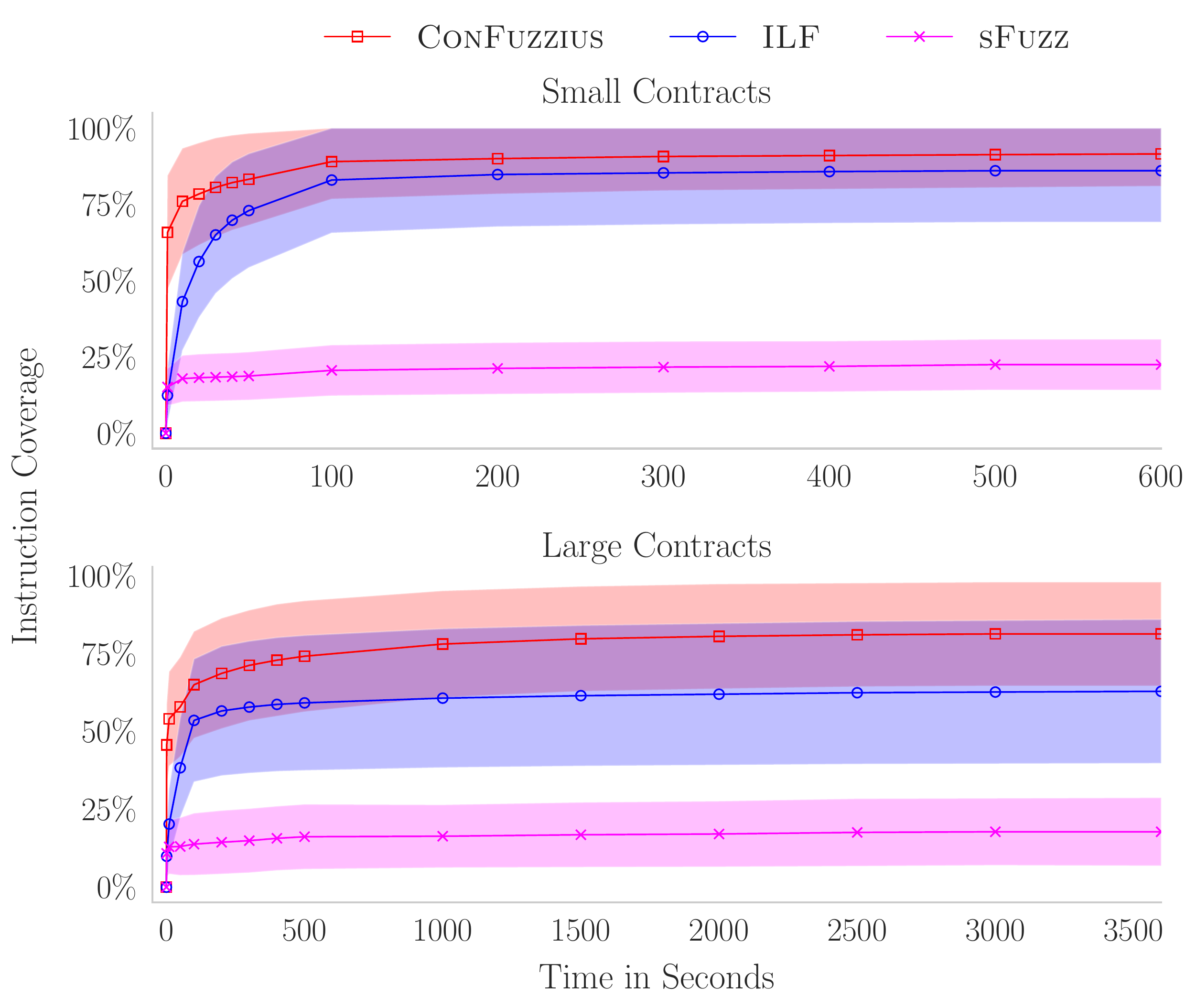}
\vspace*{-5mm}
\caption{Overall instruction coverage of \confuzzius{}, \textsc{ILF} and \textsc{sFuzz} over time.}
\label{fig:instruction_coverage}
\end{figure}

\figurename~\ref{fig:code_coverage} depicts the overall instruction coverage (\eg the average of all contract runs) of \confuzzius{} and other security tools on the clusters of small and large contracts.
\confuzzius{} achieves the highest coverage
on the small and large contracts,  with 91\% and 81\%, respectively.
As expected, every tool struggles with the larger contracts. We see that the overall coverage is less for the larger contracts than for the smaller contracts. However, while the difference is roughly 10\% for \confuzzius{}, symbolic execution tools such as \textsc{Mythril} have a difference of 31\%.
\figurename~\ref{fig:instruction_coverage} compares the overall instruction coverage of \confuzzius{} and the two other fuzzers, \textsc{ILF} and \textsc{sFuzz}, over time.
We solely plotted the instruction coverage for these three tools as we do not have coverage information over time for symbolic execution tools.
\confuzzius{} not only consistently outperforms \textsc{ILF} and \textsc{sFuzz}, but it also achieves more code coverage in a shorter time.
On the small contracts, \confuzzius{} achieves after 1 second 66\% instruction coverage, whereas \textsc{ILF} and \textsc{sFuzz} achieve solely 12\% and 15\%, respectively.
On the large contracts, \confuzzius{} achieves after 1 second
46\% instruction coverage, whereas \textsc{ILF} and \textsc{sFuzz} achieve only 10\% and 11\%, respectively.

\subsection{Vulnerability Detection}

\begin{table}[t]
\addtolength{\tabcolsep}{-3.2pt}
\centering
\caption{True positives and false positives detected by each tool per vulnerability type.}
\begin{tabular}{| l | c c c c c c c c c c | c |}
\hline
    & 
    \multicolumn{10}{c|}{\textbf{Vulnerabilities}} & \\
    \textbf{Toolname} & 
    \multicolumn{1}{c}{\textbf{AF}} &
    \multicolumn{1}{c}{\textbf{IO}} &
    \multicolumn{1}{c}{\textbf{RE}} & 
    \multicolumn{1}{c}{\textbf{TD}} &
    \multicolumn{1}{c}{\textbf{BD}} &
    \multicolumn{1}{c}{\textbf{UE}} &
    \multicolumn{1}{c}{\textbf{UD}} &
    \multicolumn{1}{c}{\textbf{LE}} &
    \multicolumn{1}{c}{\textbf{LO}} &
    \multicolumn{1}{c|}{\textbf{US}} & \textbf{Total} \\
\hline \hline
\textsc{Oyente} & 6/6 & 12/4 & 8/0 & 2/0 & 0/0 & - & - & - & - & 0/0            & 28 \\
\textsc{Mythril} & 7/3 & 18/5 & 10/0 & 0/0 & 3/0 & 24/0 & 0/0 & 4/0 & - & 2/0   & 68 \\
\textsc{M-Pro} & 7/3 & 18/5 & 10/0 & 0/0 & 3/0 & 24/0 & 0/0 & 4/0 & - & 2/0     & 68 \\
\textsc{ILF} & - & - & - & - & 0/0 & 10/0 & 1/2 & 4/0 & 5/0 & 3/0               & 23 \\
\textsc{sFuzz} & - & 12/0 & 7/0 & - & 1/0 & 21/0 & 1/2 & - & 0/0 & -            & 42 \\
\confuzzius{} & \textbf{10/0} & \textbf{18/0} & \textbf{10/0} & \textbf{2/0} & \textbf{7/0} & \textbf{46/0} & \textbf{1/0} & \textbf{4/0} & \textbf{5/0} & \textbf{3/0} & \textbf{106} \\
\hline \hline
Total Unique & 14 & 19 & 11 & 4 & 7 & 75 & 1 & 9 & 5 & 3 & 148\\ \hline
\end{tabular}
\label{tbl:vulnerabilities}
\end{table}

Table~\ref{tbl:vulnerabilities} summarizes the vulnerabilities detected by each tool for each of the 10 categories on the extended curated dataset. Each entry shows the number of true positives (left-hand side) and false positives (right-hand side). For example, \textsc{Oyente} reported for assertion failure (AF), 6 true positives and 6 false positives (\ie 6/6), whereas \textsc{Mythril} reported for assertion failure (AF), 7 true positives and 3 false positives (\ie 7/3). 
Overall, we see that \confuzzius{} detected the most number of vulnerabilities, namely 106 out of 148 vulnerabilities, that is roughly 71\% of all the vulnerabilities. \textsc{ILF} detected the least number of vulnerabilities, with 23 out of 148.
In the following, we discuss the results obtained for each category.

\noindent
\newline
\textbf{Assertion Failure (AF).}
\confuzzius{} detects more assertion failures than the other tools, namely 10 out of 14, and does not report any false positives. Both, \textsc{Mythril} and \textsc{M-Pro}, report 7 assertion failures and 3 false positives. \textsc{Oyente} reports 6 true positives and 6 false positives. 
Our manual investigation reveals that they over-approximate the satisfiability of execution paths due to incorrect modeling.
For example, \textsc{Oyente} reports in \figurename~\ref{fig:af} an assertion failure at line 8. However, this is not possible because {\small\texttt{param}} is set at the constructor and is checked to be always larger than zero.

\begin{figure}[H]
\begin{lstlisting}[language=Solidity]
contract AssertMultiTx1 {
    uint256 private param;
    constructor(uint256 _param) {
        require(_param > 0);
        param = _param;
    }
    function run() {
        assert(param > 0);
    }
}
\end{lstlisting}
\caption{False positive reported by \textsc{Oyente} on an assertion failure.}
\label{fig:af}
\end{figure}

\noindent
\textbf{Integer Overflows (IO).}
\confuzzius{} reports the same number of integer overflows as \textsc{Mythril} and \textsc{M-Pro}, namely 18 out of 19. However, \textsc{Mythril} and \textsc{M-Pro} also report 5 false positives, whereas \confuzzius{} does not report any. 
For example, \textsc{Mythril} reports in \figurename~\ref{fig:io} an integer overflow at line 6. However, there is no possibility to initialize {\small\texttt{balanceOf}}, therefore an overflow can never occur because the {\small\textbf{\texttt{\textcolor{blue}{require}}}} statement at line 4 will never be satisfied.  

\begin{figure}[H]
\begin{lstlisting}[language=Solidity]
contract IntegerOverflowAdd {
  mapping (address => uint256) balanceOf;
  function transfer(address _to, uint256 _value) {
    require(balanceOf[msg.sender] >= _value);
    balanceOf[msg.sender] -= _value;
    balanceOf[_to] += _value;
  }
}
\end{lstlisting}
\caption{False positive reported by \textsc{Mythril} on an integer overflow.}
\label{fig:io}
\end{figure}

\noindent
\textbf{Reentrancy (RE).}
\confuzzius{}, \textsc{Mythril} and \textsc{M-Pro} detect the same number of reentrancy vulnerabilities, namely 10 out of 11. \textsc{Oyente} and \textsc{sFuzz} detect 8 and 7, respectively.

\noindent
\newline
\textbf{Transaction Order Dependency (TD).}
\confuzzius{} and \textsc{Oyente} detect 2 contracts vulnerable to transaction order dependency. Both, \textsc{Mythril} and \textsc{M-Pro} do not detect any of the 4 contracts to be vulnerable to transaction order dependency.

\noindent
\newline
\textbf{Block Dependency (BD).}
\confuzzius{} is the only tool capable of detecting all the 7 block dependencies. 
Our manual investigation reveals that \confuzzius{} is capable of detecting more block dependencies because of its environmental modeling, which allows \confuzzius{} to fuzz block information and therefore, \confuzzius{} can detect more calls that are dependent on block information.

\noindent
\textbf{Unhandled Exception (UE).}
\confuzzius{} reports the largest number of unhandled exceptions, namely 46 out of 75. \textsc{Mythril} and \textsc{M-Pro} report 24 unhandled exceptions. \textsc{ILF} and \textsc{sFuzz} report 10 and 21 unhandled exceptions, respectively.
Similar to block dependency, \confuzzius{} detects more unhandled exceptions because it models call return values as environmental information that can be fuzzed. This allows \confuzzius{} to simulate exceptions and check if they are handled.

\noindent
\newline
\textbf{Unsafe Delegatecall (UD).} 
\confuzzius{} is the only tool that detects unsafe delegatecall without false positives. Both, \textsc{Mythril} and \textsc{M-Pro} were not able to detect any unsafe delegatecalls.
\textsc{ILF} and \textsc{sFuzz} detect an unsafe delegatecall, but also report 2 false positives. 
For example, in \figurename~\ref{fig:ud} \textsc{ILF} reports an unsafe delegatecall at line 13. However, the variable  {\small\texttt{callee}} can only be changed by the owner and the delegatecall is therefore safe.

\begin{figure}[H]
\begin{lstlisting}[language=Solidity]
contract Proxy {
  address callee;
  address owner;
  constructor() {
  	callee = address(0x0);
    owner = msg.sender;
  }
  function setCallee(address newCallee) {
    require(msg.sender == owner);
    callee = newCallee;
  }
  function forward(bytes _data) {
    require(callee.delegatecall(_data));
  }
}
\end{lstlisting}
\caption{False positive reported by \textsc{ILF} on an usafe delegatecall.}
\label{fig:ud}
\end{figure}

\noindent
\textbf{Leaking Ether (LE).}
\confuzzius{}, \textsc{Mythril}, \textsc{M-Pro} and \textsc{ILF} detect the same number of ether leaking vulnerabilities, namely 4 out of 9. None of the tools report false positives.

\noindent
\newline
\textbf{Locking Ether (LO).}
Both, \confuzzius{} and \textsc{ILF} detect all ether locking vulnerabilities.
\textsc{sFuzz}, on the other hand, does not detect any of the vulnerabilities.

\noindent
\newline
\textbf{Unprotected Selfdestruct (US).}
Both, \confuzzius{} and \textsc{ILF} detect all the unprotected selfdestruct vulnerabilities. 
\textsc{Oyente} does not detect any of the vulnerabilities and \textsc{Mythril} as well as \textsc{M-Pro}, detect 2 of the 3 unprotected selfdestruct vulnerabilities.

\subsection{Component Evaluation}

\begin{figure}
\centering
\vspace*{-3mm}
\includegraphics[width=0.49\textwidth]{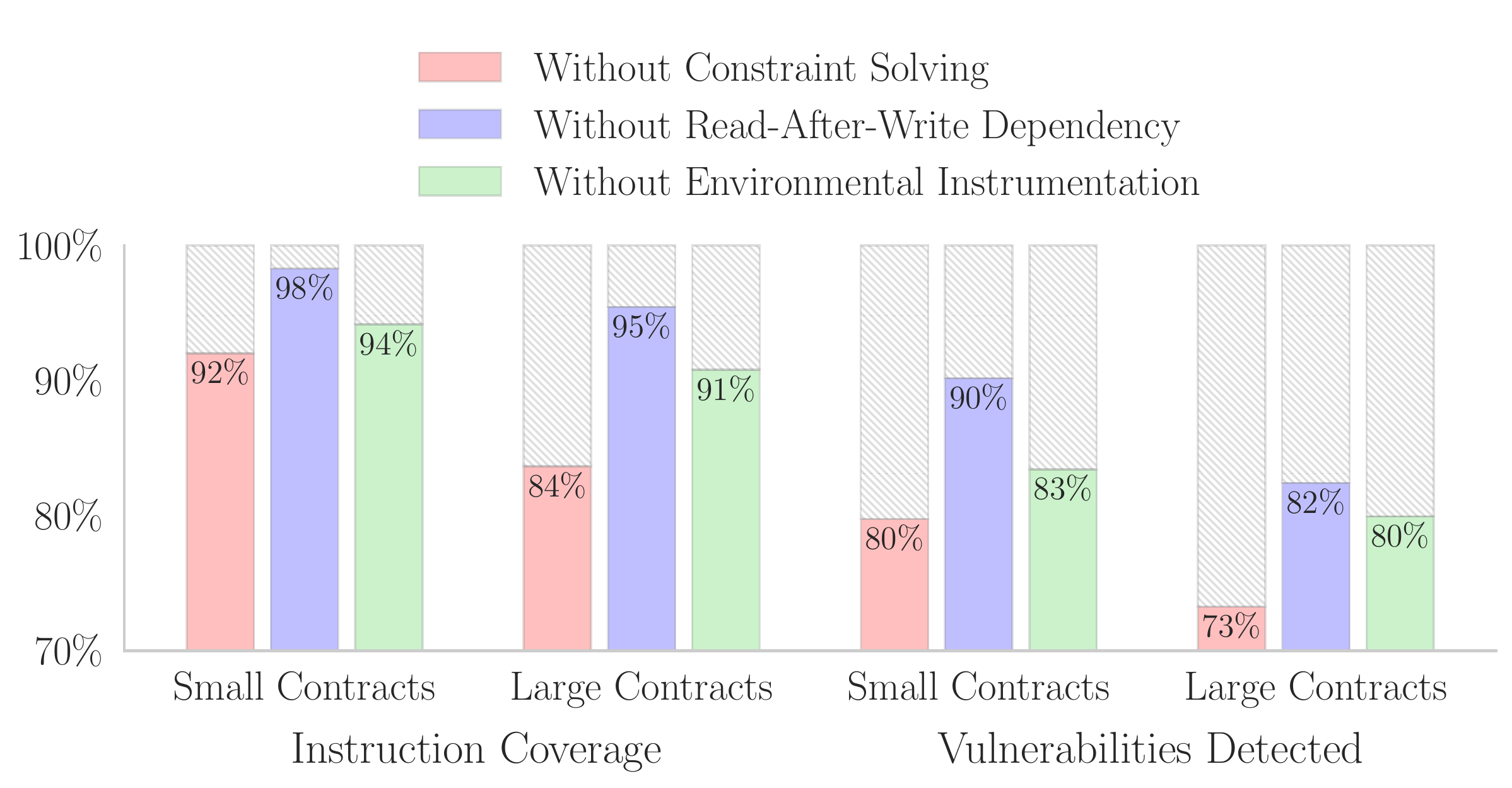}
\vspace*{-5mm}
\caption{Comparison of overall instruction coverage and vulnerabilities detected between \confuzzius{}'s individual components.}
\label{fig:components_code_coverage}
\end{figure}

In the following, we evaluate the importance of \confuzzius{}'s three main components: 1) \emph{constraint solving}, 2) \emph{read-after-write dependency analysis} and 3) \emph{environmental instrumentation}.
We randomly selected 100 contracts from each cluster. 
We then performed three experiments for each contract, where we deactivated a different component for each experiment.
For example, in the "Without Constraint Solving" experiment, we disabled constraint solving but kept RAW dependency analysis, and environmental instrumentation enabled.
This means that inputs will be generated following a random uniform distribution, and the constraint solver will not produce inputs.
In the "Without Read-After-Write Dependency" experiment, we disabled RAW dependency analysis but kept constraint solving and environmental instrumentation enabled. 
This means that transaction sequences will not be combined following RAW dependencies, but that they will be combined at random, following a uniform distribution.
Finally, in the "Without Environmental Instrumentation" experiment, we disabled environmental instrumentation but kept constraint solving and RAW dependency analysis enabled.
This means that environmental information such as block timestamps or call return values will not be fuzzed.
We repeated each experiment 10 times and run the experiments on the small contracts for 10 minutes and the large contracts for 1 hour.
In \figurename~\ref{fig:components_code_coverage}, each bar depicts the percentage of the achieved results compared to the results when all three components were enabled (\ie the gray bar in the back).
We see that each component is an added value for \confuzzius{}, which means that they are all essential to \confuzzius{}'s performance.
However, we can see that generating meaningful inputs via constraint solving plays an essential part in achieving broad code coverage and detecting more bugs. 
Also, environmental instrumentation seems to help achieve code coverage and detect bugs. 
Nevertheless, our novel method that leverages RAW dependency analysis to create meaningful sequences of inputs at runtime can provide 2\% more code coverage on small contracts and 5\% on large contracts.
Further, it allows the detectors to find 10\% and 18\% more vulnerabilities in small and large contracts, respectively.

\section{Related Work}
\label{sec:related_work}

Since its introduction by Miller \etal \cite{miller1990empirical}, fuzzing has been applied to many different domains and targets.

\noindent
\newline
\textbf{Software Fuzzing.}
American Fuzzy Loop (AFL)~\cite{afl2016} is one of the most widespread fuzzers and it is based on evolutionary fuzzing and exploits execution data to guide the generation/mutation of fuzzed inputs.
Besides AFL and its offsprings\cite{bohme2017coverage,bohme2017directed}, other fuzzers also use evolutionary approaches to generate test inputs automatically\cite{rawat2017vuzzer,iannillo2017chizpurfle}. 
On the other hand, KLEE~\cite{cadar2008klee} and SAGE~\cite{godefroid2008automated} are white-box fuzzers, that execute code in a controlled environment.
Driller~\cite{stephens2016driller} is a hybrid fuzzer that leverages selective concolic execution in a complementary manner.
Symbolic execution based fuzzers produce meaningful inputs but tend to be slow\cite{chen2018angora,peng2018t,chen2019matryoshka,cho2019intriguer}.
Fuzzers such as LibFuzzer~\cite{serebryany2016continuous}, FuzzGen~\cite{ispoglou2020fuzzgen} and FUDGE~\cite{babic2019fudge} focus on fuzzing libraries, which cannot run as standalone programs, but instead are invoked by other programs.

\noindent
\newline
\textbf{Smart Contract Fuzzing.} \textsc{ContractFuzzer}  \cite{jiang2018contractfuzzer} generates inputs based on a list of input seeds.
While \textsc{ContractFuzzer} deploys an entire custom testnet to fuzz transactions, \confuzzius{} is more efficient and solely emulates the EVM.
Moreover, \confuzzius{} does not rely on user-provided input seeds but instead analyzes the execution traces and feeds constraints related to the execution to a constraint solver in order to generate new values specific to the contract under test.
\textsc{Echidna}~\cite{echidna2020} is a property-based testing tool for smart contracts that leverages grammar-based fuzzing.
\textsc{Echidna} requires user-defined predicates in the form of Solidity assertions and does not automatically check for vulnerabilities.
\textsc{Harvey} \cite{wustholz2019harvey} predicts new inputs based on instruction-granularity cost metrics.
In contrast, \confuzzius{} exploits lightweight symbolic execution when the population fitness does not increase (see Section~\ref{sec:execution_trace_analyzer}).
Further, \textsc{Harvey} fuzzes transaction sequences in a targeted and demand-driven way, assisted by an aggressive mode that directly fuzzes the persistent state of a smart contract.
Instead, \confuzzius{} relies on the read-after-write dependencies to guide the selection and crossover operators to create meaningful transaction sequences efficiently (see Section~\ref{sec:evolutionary_fuzzing_engine}).
\textsc{ILF} \cite{he2019ilf} is based on imitation learning, which requires a learning phase prior to fuzzing. 
\textsc{ILF} consists of a neural network that is trained on transactions obtained by running a symbolic execution expert over a broad set of contracts.
\confuzzius{} does not have the overhead of the learning phase and uses on-demand constraint solving while actively fuzzing the target. 
Moreover, \textsc{ILF} is limited to the knowledge that it learned during the learning phase, meaning that \textsc{ILF} has issues in getting past program conditions that require inputs that were not part of the learning dataset. \confuzzius{} does not have this issue as it learns tailored inputs per target while fuzzing.
\textsc{sFuzz} \cite{nguyen2020sfuzz} is an AFL based smart contract fuzzer, whereas \textsc{EthPloit} \cite{zhang2020ethploit} is a fuzzing based smart contract exploit generator.
Both \textsc{sFuzz} and \textsc{EthPloit} have been developed concurrently and independently of \confuzzius{}.
While \textsc{sFuzz} follows a random strategy to create transaction sequences, \textsc{EthPloit} uses static taint analysis on state variables to create meaningful transaction sequences. 
However, static taint analysis has the disadvantage of being imprecise and analyzing parts that are not executable.
Despite \textsc{sFuzz} using a genetic algorithm as \confuzzius{}, it employs a different encoding of individuals.
\textsc{sFuzz} only models block number and timestamp as environmental information.
\textsc{EthPloit}, on the other hand, instruments the EVM in a similar way to \confuzzius{}. 
However, \textsc{EthPloit} does not fuzz the size of external code or contract call return values.

\noindent
\newline
\textbf{Smart Contract Symbolic Execution.}
Apart from fuzzing, several other tools based on symbolic execution were proposed to assess the security of smart contracts \cite{Luu2016}\cite{nikolic2018finding}\cite{mueller2018}\cite{torres2018osiris}\cite{torres2019art}\cite{torres2019art}\cite{krupp2018teether}\cite{frank2020ethbmc}.
\textsc{MPro} \cite{zhang2019mpro} combines symbolic execution and data dependency analysis to deal with the scalability issues that symbolic execution tools face when trying to handle the statefulness of smart contracts.
\textsc{MPro} has been developed concurrently and independently from \confuzzius{}.
There are two significant differences in our approach. 
First, \textsc{MPro} retrieves data dependencies using static analysis and therefore requires source code, whereas \confuzzius{} tries to infer data dependencies from bytecode. 
Second, \textsc{MPro} works in two separate steps, first, it infers data dependencies via static analysis, and then it applies symbolic execution. \confuzzius{}, on the other hand, applies a dynamic approach and infers data dependencies while fuzzing.
\textsc{EthRacer} \cite{kolluri2018exploiting} uses a hybrid approach with a converse strategy by primarily using symbolic execution to test a smart contract and using fuzzing only for producing combinations of transactions to detect vulnerabilities such as transaction order dependency. \confuzzius's fuzzing strategy, compared to \textsc{EthRacer}, is not entirely random but based on read-after-write dependencies, yielding faster and more efficient combinations.

\noindent
\newline
\textbf{Smart Contract Static Analysis.}
Besides symbolic execution and fuzzing, other works based on static analysis were proposed to detect smart contract vulnerabilities.
\textsc{Zeus} \cite{kalra2018zeus} is a framework for automated verification of smart contracts using abstract interpretation and model checking.
\textsc{Securify} \cite{tsankov2018securify} uses static analysis based on a contract's dependency graph to extract semantic information about the program bytecode and then checks for violations of safety patterns. 
Similarly, \textsc{Vandal} \cite{brent2018vandal} is a framework designed to convert EVM bytecode into semantic logic relations in Datalog, which can then be queried for vulnerabilities.

\section{Conclusion}
\label{sec:conclusion}

We presented \confuzzius{}, the first hybrid fuzzer for smart contracts. It tackles the three main challenges of smart contract testing: \emph{input generation}, \emph{stateful exploration}, and \emph{environmental dependencies}.
We solved the first challenge by combining evolutionary fuzzing with constraint solving to generate meaningful inputs. 
The second challenge is solved by leveraging data dependency analysis across state variables to generate purposeful transaction sequences.
Finally, we solved the third challenge by modeling block related information (e.g. block number) and contract related information (e.g. call return values) as fuzzable inputs.
We run \confuzzius{} and other state-of-the-art fuzzers and symbolic execution tools for smart contracts against a curated dataset of 128 contracts and a dataset of 21K real-world smart contracts.
Our results show that our hybrid approach not only detects more bugs than existing state-of-the-art tools (up to 23\%), but also that \confuzzius{} outperforms these tools in terms of code coverage (up to 69\%) and that data dependency analysis can boost the detection of bugs (up to 18\%).

\section*{Acknowledgment}

We would like to thank the anonymous reviewers for their valuable comments and feedback. This work was partly supported by the Luxembourg National Research Fund (FNR) under grant 13192291 and is part of a project that has received funding from the European Union’s Horizon 2020 research and innovation programme under grant agreement No 830927.

\bibliographystyle{IEEEtranS}
\balance
\bibliography{references}

\appendices

\section{Vulnerability Detectors}

\noindent
We elaborate on the implementation details of our 10 vulnerability detectors below.

\noindent
\newline
\textbf{Assertion Failure.} 
We detect an assertion failure by checking if the execution trace contains an 
\code{ASSERTFAIL} or \code{INVALID} instruction.

\noindent
\newline
\textbf{Integer Overflow.} 
Detecting integer overflows is not trivial, since not every overflow is considered harmful. 
Integer overflows may also be introduced by the compiler for optimization purposes.
Therefore, we only consider an overflow as harmful, if it modifies the state of the smart contract, \ie if the result of the computation is written to storage or is used to send funds.
We follow the approach by Torres \etal \cite{torres2018osiris} and start by analyzing if the execution trace contains an \code{ADD}, \code{MUL} or \code{SUB} instruction. 
We then extract the operands from the stack and use these to compute the result of the arithmetic operation. 
Afterwards, we check if our result is equivalent to the result that has been pushed onto the stack.
If they are not the same, we know that an integer overflow has occurred and we keep track of the overflow by tainting the result of the computation.
We report an integer overflow if the tainted result flows into an \code{SSTORE} instruction or a \code{CALL} instruction, as these instructions will result in updating the blockchain state.

\noindent
\newline
\textbf{Reentrancy.} 
A reentrancy occurs whenever a contract calls another contract, and that contract calls back the original contract.
We detect reentrancy by first checking if the execution trace contains a \code{CALL} instruction whose gas value is larger than 2,300 units and where the amount of funds to be transferred is a symbolic value or a concrete value that is larger than zero.
Finally, we report a reentrancy if we find an \code{SLOAD} instruction that occurs before the \code{CALL} instruction and an \code{SSTORE} instruction that occurs after the \code{CALL} instruction, and which shares the same storage location as the \code{SLOAD} instruction.

\noindent
\newline
\textbf{Transaction Order Dependency.} 
We detect transaction order dependency by checking if there are two execution traces with different senders, where the first execution trace writes to the same storage location from which the second execution trace reads.

\noindent
\newline
\textbf{Block Dependency.}
We detect a block dependency by checking if the execution trace contains either a \code{CREATE}, \code{CALL}, \code{DELEGATECALL}, or \code{SELFDESTRUCT} instruction, that is either control-flow or data dependent on a \code{BLOCKHASH}, \code{COINBASE}, \code{TIMESTAMP}, \code{NUMBER}, \code{DIFFICULTY}, or \code{GASLIMIT} instruction.

\noindent
\newline
\textbf{Unhandled Exception.}
We detect unhandled exceptions by first checking if the execution trace contains a \code{CALL} instruction that pushes to the stack the value $1$ as a result of the call. A value of $1$ means that an error occurred during the call (\ie an exception).
Afterwards, we check if the result of the call flows into a \code{JUMPI} instruction. If the result does not flow into a \code{JUMPI} instruction until the end of the execution trace, then this means that the exception of the call was not handled and we report an unhandled exception.

\noindent
\newline
\textbf{Unsafe Delegatecall.}
We detect an unsafe delegate call by checking if there is an execution trace that contains a \code{DELEGATECALL} instruction and terminates with a \code{STOP} instruction, but whose sender is an attacker address.
Attacker and benign user addresses are generated at the start by the fuzzer.

\noindent
\newline
\textbf{Leaking Ether.}
We detect the leaking of ether by checking if the execution trace contains a \code{CALL} instruction, whose recipient is an attacker address that has never sent ether to the contract in a previous transaction and has never been passed as a parameter in a function by an address that is not an attacker.

\noindent
\newline
\textbf{Locking Ether.}
We detect the locking of ether by checking if a contract can receive ether but cannot send out ether. 
To check if a contract cannot send ether, we check if the runtime bytecode of the contract does not contain any \code{CREATE}, \code{CALL}, \code{DELEGATECALL}, or \code{SELFDESTRUCT} instruction.
To check if a contact can receive ether, we check if the execution trace has a transaction value larger than 0 and terminates with a \code{STOP} instruction.

\noindent
\newline
\textbf{Unprotected Selfdestruct.}
Similar to the leaking ether or unsafe delegatecall vulnerability detectors, this detector relies on attacker accounts.
We detect an unprotected selfdestruct by checking if the execution trace contains a \code{SELFDESTRUCT} instruction where the sender of the transaction is an attacker and its address has not been previously passed as an argument by a benign user.

\end{document}